\documentstyle[12pt,epsf]{article}
 
\begin{document}
 
\begin{titlepage}
 
\hspace*{\fill}\parbox[t]{2.8cm}{DESY 95-249 \\ December 1995}
 
\vspace*{1cm}
 
\begin{center}
\large\bf
Real next-to-leading corrections to the multigluon amplitudes in the
helicity formalism
\end{center}
 
\vspace*{0.5cm}
 
\begin{center}
Vittorio Del Duca \\
Deutsches Elektronen-Synchrotron \\
DESY, D-22603 Hamburg , GERMANY\footnote{After January 1, 1996:
Dept. of Physics and Astronomy, University of Edinburgh, Edinburgh EH9 3JZ,
Scotland}
\end{center}
 
\vspace*{0.5cm}
 
\begin{center}
\bf Abstract
\end{center}
 
\noindent

Using the helicity formalism, we compute the corrections to the tree-level 
multigluon amplitudes in the high-energy limit, induced by the corrections 
to the multi-Regge kinematics, and we show that they coincide with the
corresponding Fadin-Lipatov amplitudes at fixed helicities.

\end{titlepage}
 
\baselineskip=0.8cm
 
\section{Introduction}
\label{sec:one}

The Fadin-Kuraev-Lipatov (FKL) multigluon amplitudes \cite{lip}, \cite{FKL} 
are the building blocks of the Balitsky-Fadin-Kuraev-Lipatov (BFKL) 
theory \cite{BFKL}, which describes the dynamics of a short-distance 
strong-interaction process in the limit of high squared parton 
center-of-mass energy $\hat s$ and fixed
momentum transfer $\hat t$. In the BFKL theory, the leading logarithmic
contributions, in $\ln(\hat s/\hat t)$, to the scattering amplitude with
exchange of a color-singlet two-gluon ladder in the crossed channel are
computed.
At $\hat t=0$ the amplitude is related via the $\hat s$-channel unitarity
to the total parton cross section with exchange of a one-gluon ladder. 

The FKL multigluon amplitudes are computed in the multi-Regge kinematics,
which requires that the final-state gluons are strongly ordered in
rapidity and have comparable transverse momentum. They
assume a simpler analytic form when the helicities of the incoming 
and the outgoing gluons are explicitly fixed \cite{lipat} and 
\cite{ptlip}. In addition, in ref.~\cite{ptlip}
it was shown that at the tree level, and for the helicity configurations 
they have in common, the FKL amplitudes coincide with the Parke-Taylor (PT) 
amplitudes \cite{pt} computed in the multi-Regge kinematics.

The BFKL theory has several phenomenological applications, among which
to the evolution of the $F_2$ structure function in deeply
inelastic scattering in the small $x_{bj}$ regime \cite{bcm}, and to the 
behavior of the dijet production cross section at large rapidity intervals
in hadron-hadron collisions \cite{mn}. The phenomenology, however,
is limited from large theoretical uncertainties, due to the lack of
knowledge of the next-to-leading logarithmic corrections to the BFKL theory.
Accordingly, real next-to-leading corrections \cite{fl}, \cite{fiore},
\cite{progress} induced 
by the corrections to the multi-Regge kinematics, and virtual  
next-to-leading-logarithmic corrections \cite{fiore}, \cite{ff} to the FKL 
amplitudes have been computed. The real next-to-leading corrections, once 
integrated over the phase space, will yield the real 
next-to-leading-logarithmic corrections to the BFKL equation.

The corrections to the tree-level FKL amplitudes \cite{fl}, \cite{fiore},
\cite{progress}
arise from the kinematical regions in which two gluons or a quark-antiquark 
pair are produced with
likewise rapidity, either at the ends of or along the ladder, which we term
the forward-rapidity and the central-rapidity regions, respectively.
The goal of this paper is to reexamine the purely gluonic corrections \cite{fl}
to the tree-level FKL amplitudes from the standpoint of the helicity formalism.

In sect.~\ref{sec:due} we summarize the contents of ref.~\cite{ptlip},
namely the computation of the FKL amplitudes in the helicity formalism
and their equivalence to the PT amplitudes in the multi-Regge kinematics.
In sect.~\ref{sec:tre} we compute the next-to-leading corrections to
the FKL amplitudes in the forward-rapidity region: $a)$ by using the helicity
formalism, i.e. by specifying the PT amplitudes to the 
kinematics with two gluons produced with likewise rapidity
in a forward-rapidity region (sect.~\ref{sec:treone}); $b)$ by fixing the
helicities in the Fadin-Lipatov amplitudes \cite{fl} 
(sect.~\ref{sec:tredue}). The amplitudes, computed in the two
different ways, coincide and have a fairly simple analytic form at fixed 
helicities.
As a bookkeeping device to list the leading color configurations we 
have found useful the two-sided lego-plot picture \cite{bj} which
we used extensively in the multi-Regge kinematics \cite{ptlip} and \cite{vd}.
In sect.~\ref{sec:four} we consider the next-to-leading corrections to
the FKL amplitudes in the central-rapidity region. In the helicity
formalism, the PT amplitudes allow us to compute the configurations
with the two gluons in the central-rapidity region produced with
equal helicity (sect.~\ref{sec:fourone}), however to analyse the case in
which the two central gluons have opposite helicities we must
consider the amplitudes with 3 negative-helicity gluons \cite{bg}, \cite{mpz}
(sect.~\ref{sec:fourbis}). In sect.~\ref{sec:fourdue} we fix the
helicities in the Fadin-Lipatov amplitudes and show that they
coincide with the respective amplitudes in the helicity formalism
(sect.~\ref{sec:fourone} and \ref{sec:fourbis}). 
We note, though, that both for the corrections in the
forward-rapidity (sect.~\ref{sec:tre}) and in the central-rapidity
(sect.~\ref{sec:four}) regions, the algebra involved
is simpler starting from the helicity amplitudes than by fixing the
helicities in the Fadin-Lipatov amplitudes. In sect.~\ref{sec:conc} we
summarize the results of this paper and present our conclusions.

\section{The multigluon amplitudes in the multi-Regge kinematics in the
helicity formalism}
\label{sec:due}

\begin{figure}[hbt]
\vspace*{-4.5cm}
\hspace*{2.0cm}
\epsfxsize=15cm \epsfbox{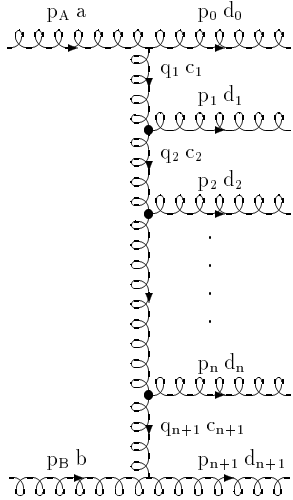}
\vspace*{-10.0cm}
\caption{Multigluon amplitude in multiregge kinematics at the tree level. 
The blobs remind that non-local effective Lipatov vertices are used for the 
gluon emissions along the ladder.}
\label{fig:fkl}
\end{figure}

We consider the production of $n+2$ gluons of momentum $p_i$, with 
$i=0,...,n+1$ and $n\ge 0$, in the scattering between two gluons of momenta 
$p_A$ and $p_B$, and we assume that the produced gluons fulfill
the multi-Regge kinematics, i.e. we require that the gluons
are strongly ordered in rapidity $y$ and have comparable transverse momentum,
\begin{equation}
y_0 \gg y_1 \gg ...\gg y_{n+1};\qquad |p_{i\perp}|\simeq|p_{\perp}|\, 
.\label{mreg}
\end{equation}
The tree-level amplitude for the production of $n+2$ gluons in the
multiregge kinematics has been computed in ref.\cite{FKL} 
(Fig.\ref{fig:fkl}), 
\begin{eqnarray}
i M^{ad_0...d_{n+1}b}_{\nu_A\nu_0...\nu_{n+1}\nu_B} &=& 2i\, {\hat s}
\left(i g\, f^{ad_0c_1}\, \Gamma^{\mu_A\,\mu_0}\right)\, 
\epsilon_{\mu_A}^{\nu_A*}(p_A) \epsilon_{\mu_0}^{\nu_0}(p_0)\, {1\over\hat t_1}
\nonumber\\ &\times& \left(i g\, f^{c_1d_1c_2}\, C^{\mu_1}(q_1,q_2)\right)\,
\epsilon_{\mu_1}^{\nu_1}(p_1)\, {1\over \hat t_2} \nonumber\\ &\times&
\label{ntree}\\ &\times&\nonumber\\ &\times& \left(i g\, f^{c_nd_nc_{n+1}}\, 
C^{\mu_n}(q_n,q_{n+1})\right)\, \epsilon_{\mu_n}^{\nu_n}(p_n)\,
{1\over \hat t_{n+1}} \nonumber\\ &\times& \left(i g\, f^{bd_{n+1}c_{n+1}}\, 
\Gamma^{\mu_b\,\mu_{n+1}}\right)\, \epsilon_{\mu_B}^{\nu_B*}(p_B) 
\epsilon_{\mu_{n+1}}^{\nu_{n+1}}(p_{n+1})\, ,\nonumber
\end{eqnarray}
where $a,d_0,..., d_{n+1},b$, and $\epsilon_A, \epsilon_0,..., \epsilon_B$ 
are respectively the colors and the polarizations of the gluons, the 
$\nu$'s are the helicities, the $q$'s are the momenta of the gluons 
exchanged in the $\hat t$ channel, and $\hat t_i = q_i^2 \simeq 
-|q_{i\perp}|^2$. The $\Gamma$-tensors \cite{FKL} are,
\begin{eqnarray}
\Gamma^{\mu_A\mu_0}(p_A, p_0, p_B) &=& g^{\mu_A\mu_0} - 
{p_A^{\mu_0} p_B^{\mu_A} \over p_A\cdot p_B} - {p_B^{\mu_0} p_0^{\mu_A} 
\over p_0\cdot p_B} + p_B^{\mu_0} p_B^{\mu_A}\, {p_A\cdot p_0\over p_A\cdot 
p_B\, p_0\cdot p_B}\, ,\label{gamm}\\
\Gamma^{\mu_B\mu_{n+1}}(p_B, p_{n+1}, p_A) &=& g^{\mu_B\mu_{n+1}} - 
{p_A^{\mu_B} p_B^{\mu_{n+1}}\over p_A\cdot p_B} - 
{p_A^{\mu_{n+1}} p_{n+1}^{\mu_B} \over p_A\cdot p_{n+1}} +
p_A^{\mu_{n+1}} p_A^{\mu_B}\, {p_B\cdot p_{n+1} \over p_A\cdot p_B\,
p_A\cdot p_{n+1}}\, ,\nonumber
\end{eqnarray}
and the Lipatov vertex \cite{lip} is 
\begin{equation}
C^{\mu}(q_i,q_{i+1}) = \left[(q_i+q_{i+1})^{\mu}_{\perp}\, -\, 
\left({\hat s_{Ai}\over\hat s}\,+\,2{\hat t_{i+1}\over\hat s_{Bi}}\right) 
p_B^{\mu}\, + \left({\hat s_{Bi}\over\hat s}\,+\,2{\hat t_i\over\hat s_{Ai}}
\right) p_A^{\mu}\right]\, ,\label{nver}
\end{equation}
with $q_{i\perp}^{\mu}=(0,0;q_{i\perp})$ and the Mandelstam invariants as 
given by eq.(\ref{mrinv}) (Appendix \ref{sec:appb}). The $\Gamma$-tensors and 
the Lipatov vertex are gauge invariant,
\begin{eqnarray}
\Gamma^{\mu_A\mu_0}(p_A, p_0, p_B)\, p_0^{\mu_0} &=&
\Gamma^{\mu_A\mu_0}(p_A, p_0, p_B)\, p_A^{\mu_A} = 0 \label{gauge}\\
C(q_i,q_{i+1}) \cdot p_i &=& 0\, ,\nonumber
\end{eqnarray}
making thus the amplitude (\ref{ntree}) invariant with respect to
arbitrary gauge transformations. The $\Gamma$-tensors conserve the helicity 
at the production vertices for the first and the last gluon along the ladder.
The square of the Lipatov vertex (\ref{nver}), which forms the kernel of the
BFKL equation \cite{BFKL}, is
\begin{eqnarray}
& & C^{\mu}(q_i,q_{i+1}) C^{\mu'}(q_i,q_{i+1}) \sum_{\nu} 
\epsilon_{\mu}^{\nu}(p_i) \epsilon_{\mu'}^{\nu*}(p_i) \label{square}\\
& & = - C^{\mu}(q_i,q_{i+1}) C^{\mu'}(q_i,q_{i+1}) g_{\mu\mu'} 
= 4\, {|q_{i\perp}|^2 |q_{i+1\perp}|^2 \over |p_{i\perp}|^2}\, ,\nonumber
\end{eqnarray}
using the gauge invariance.
Thus in the square (\ref{square}) of the Lipatov vertex several cancellations
have occurred, making the square of the vertex simpler than the vertex itself.
One of the goals in using the helicity formalism, which we now turn to, is 
to achieve the simplicity of eq.(\ref{square}) already at the amplitude level.

\subsection{The Fadin-Kuraev-Lipatov amplitudes at fixed helicities}
\label{sec:dueone}

We choose the representation (\ref{hpol}) (Appendix \ref{sec:appa}) for 
the polarizations. This is equivalent to use a physical gauge. Accordingly,
we must specify a reference vector with respect to which
we compute the polarizations in eq.(\ref{ntree}), but thanks to the gauge
invariance the choice is arbitrary. The contraction of the 
Lipatov vertex with the gluon polarization in eq.(\ref{ntree}) is
\cite{lipat}, \cite{ptlip},
\begin{equation}
\epsilon^+(p_i, p_A)\cdot C(q_i,q_{i+1}) = \sqrt{2}\, {q_{i\perp}^* 
q_{i+1\perp}\over p_{i\perp}}\, .\label{verc}
\end{equation}
From eq.(\ref{exh}) (Appendix \ref{sec:appa}), the contractions of the 
helicity-conserving tensors (\ref{gamm}) with the gluon polarizations are, 
\begin{eqnarray}
\Gamma^{\mu_B\mu_{n+1}}\,\epsilon^{+*}_{\mu_B}(p_B, p_A)\, 
\epsilon^+_{\mu_{n+1}}(p_{n+1}, p_A) &=& -{p_{n+1\perp}^*\over
p_{n+1\perp}}\, ,\label{contra}\\
\Gamma^{\mu_A\mu_0}\,\epsilon^{+*}_{\mu_A}(p_A, p_B)\, \epsilon^+_{\mu_0}(p_0, 
p_B) &=& -1\, .\nonumber
\end{eqnarray}
The product of the structure constants in eq.(\ref{ntree}) may be rewritten
as a trace of nested commutators of $\lambda$ matrices, the color matrices 
in the fundamental representation of SU($N_c$)
\begin{eqnarray}
& & f^{ad_0c_1}\, f^{c_1d_1c_2}\,\cdots f^{c_nd_nc_{n+1}} f^{bd_{n+1}c_{n+1}}
\nonumber\\ & & = -2\, (-i)^{n+2}\, {\rm tr}\left(\lambda^a\,\left[
\lambda^{d_0}, \left[\lambda^{d_1},...,\left[\lambda^{d_{n+1}}, \lambda^b
\right]\right]\right]\right) \label{fpro}\\ & & =
-2\, (-i)^{n+2}\, {\rm tr}\left(\lambda^a\lambda^{d_0} \cdots 
\lambda^{d_{n+1}} \lambda^b -\sum_{j=0}^{n+1} \lambda^a\lambda^{d_0} \cdots 
\lambda^{d_{j-1}} \lambda^{d_{j+1}}\cdots \lambda^{d_{n+1}} \lambda^b 
\lambda^{d_j}\right. \nonumber\\ & & \left. + \sum_{j<k} \lambda^a
\lambda^{d_0} \cdots \lambda^{d_{j-1}} \lambda^{d_{j+1}}\cdots 
\lambda^{d_{k-1}} \lambda^{d_{k+1}}\cdots\lambda^{d_{n+1}} \lambda^b 
\lambda^{d_k} \lambda^{d_j} + \cdots\right)\, .\nonumber
\end{eqnarray}
Substituting eq.(\ref{verc}), (\ref{contra}) and (\ref{fpro}) into 
eq.(\ref{ntree}), the FKL amplitude for the configuration with all the
gluons having helicity $\nu=+$ becomes,
\begin{eqnarray}
& & i M(p_A,+; p_0,+;...; p_{n+1},+; p_B,+) \label{fklh}\\ & & =
i (-1)^{n+1}\, 2^{2+n/2}\, g^{n+2}\, {\hat s}\, {1\over \prod_{i=0}^{n+1} 
p_{i\perp}}\, {\rm tr}\left(\lambda^a\,\left[\lambda^{d_0},\left[
\lambda^{d_1},...,\left[\lambda^{d_{n+1}}, \lambda^b\right]\right]\right]
\right)\, .\nonumber
\end{eqnarray}
The configuration with all the helicities $\nu = -$ is obtained by
replacing the complex conjugates of eq.(\ref{verc}) and (\ref{contra}) into 
eq.(\ref{ntree}), which amounts to exchange $\prod_i p_{i\perp}$ with
$\prod_i p_{i\perp}^*$ in eq.(\ref{fklh}).

The calculation of the FKL amplitude for the other helicity 
configurations at the helicity-conserving vertices is obtained from the one of 
eq.(\ref{fklh}), 
by taking the suitable complex conjugates of the contractions (\ref{contra}),
\begin{eqnarray}
M(p_A,-; p_0,-;...; p_{n+1},+; p_B,+) &=& M(p_A,+; p_0,+;...; p_{n+1},+;
p_B,+)\, ,\label{fklhel}\\
M(p_A,+; p_0,+;...; p_{n+1},-; p_B,-) &=& M(p_A,-; p_0,-;...; p_{n+1},-; 
p_B,-)\, ,\label{fklhelb}\\ &=& \left({p_{n+1\perp}\over p_{n+1\perp}^*}
\right)^2\, M(p_A,+; p_0,+;...; p_{n+1},+; p_B,+)\, ,\nonumber
\end{eqnarray}
with helicities $\nu_i = +$ and $i=1,...,n$. 
Finally, changing the helicity of a gluon along the ladder by means of
eq.(\ref{verc}), the amplitude (\ref{fklh}) changes only by a phase,
\begin{eqnarray}
& & M(p_A,+; p_0,+;...; p_{j-1},+; p_j,-; p_{j+1},+;...; p_{n+1},+; p_B,+)
\label{chh}\\
& & = {p_{i\perp} q_{i\perp} q_{i+1\perp}^*\over p_{i\perp}^* q_{i\perp}^* 
q_{i+1\perp}}\, M(p_A,+; p_0,+;...; p_{n+1},+; p_B,+)\, .\nonumber
\end{eqnarray}

\subsection{The Parke-Taylor amplitudes in the multi-Regge kinematics}
\label{sec:duedue}

A tree-level multigluon amplitude in a helicity basis has
in general the form \cite{mp}
\begin{equation}
M_n = \sum_{[A,0,...,n+1,B]} {\rm tr}(\lambda^a\lambda^{d_0} \cdots
\lambda^{d_{n+1}} \lambda^b) \, m(\tilde{p}_A,\epsilon_A; p_0,\epsilon_0;...;
p_{n+1},\epsilon_{n+1}; \tilde{p}_B,\epsilon_B)\, ,\label{one}
\end{equation}
where the sum is over the noncyclic
permutations of the color orderings $[A,0,...,B]$. Considering all the momenta 
as outgoing, the PT amplitudes describe the {\sl maximally helicity-violating}
configurations $(-,-,+,...,+)$ for which the subamplitudes, \newline
$m(\tilde{p}_A,\epsilon_A; p_0,\epsilon_0;...; p_{n+1},\epsilon_{n+1}; 
\tilde{p}_B,\epsilon_B)$, invariant with
respect to tranformations between physical gauges,
assume the form \cite{pt}, \cite{mp}\footnote{Note 
that eq.(\ref{two}) differs for the $\sqrt{2}$ factors from the expression
given in ref.\cite{mp}, because we use the standard normalization of
the $\lambda$ matrices, ${\rm tr}(\lambda^a\lambda^b) = \delta^{ab}/2$.},
\begin {equation}
i m(-,-,+,...,+) = 2^{2+n/2}\, i\, g^{n+2}\, {\langle p_i p_j\rangle^4\over
\langle \tilde{p}_A p_0\rangle \cdots\langle p_{n+1} \tilde{p}_B\rangle 
\langle \tilde{p}_B \tilde{p}_A\rangle}\, ,\label{two}
\end{equation}
where $i$ and $j$ are the gluons of negative helicity, and the ordering
of the spinor products in the denominator is set by the permutation of
the color ordering $[A,0,...,B]$. The subamplitudes
(\ref{two}) are {\sl exact}, i.e. valid for arbitrary kinematics.
In eq.(\ref{two})
the representation (\ref{hpol}) (Appendix \ref{sec:appa}) for the 
gluon polarization has been used. The configurations
$(+,+,-,...,-)$ are then obtained by replacing the $\langle p k\rangle$
products with $\left[k p\right]$ products. 

\begin{figure}[htb]
\vspace{12pt}
\vskip 0cm
\epsfysize=6cm
\centerline{\epsffile{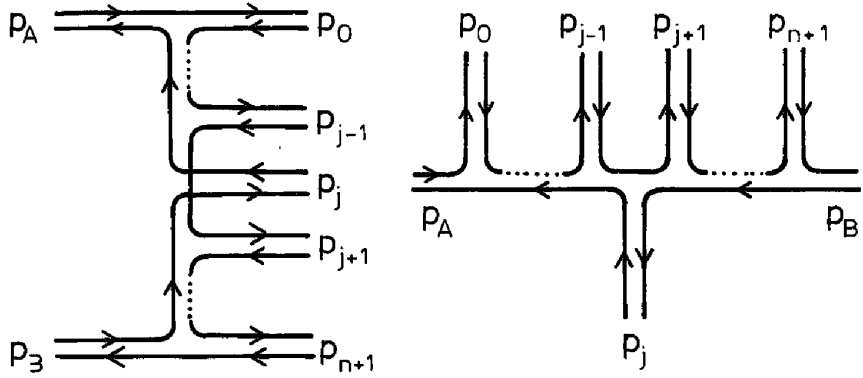}}
\vspace{18pt}
\vskip-1cm
\caption{$a)$ PT amplitude with color ordering [$A,0,...,j-1,j+1,...,n+1,B,
j$], and $b)$ its untwisted version on the two-sided lego plot.} 
\label{fig:double}
\vspace{12pt}
\end{figure}

Computing the spinor products in eq.(\ref{two}) by means of eq.(\ref{mrpro})
(Appendix \ref{sec:appb}), we find that the leading color orderings 
in the multi-Regge kinematics are 
given by the $2^{n+2}$ configurations which respect the rapidity ordering 
on the two-sided lego plot in rapidity and azimuthal angle \cite{ptlip}, 
\cite{bj}, \cite{vd}. The second side of the lego plot is obtained by
untwisting the color lines in a given color ordering. An example for the
color orderings [$A,0,...,j-1,j+1,...,n+1,B,j$], with $j=0,...,n+1$, is 
given in Fig.~\ref{fig:double}. All the leading color orderings are
the ones given by eq.(\ref{fpro}). Then for the helicity configurations 
of eq.(\ref{fklh}), (\ref{fklhel}) and (\ref{fklhelb}), the PT
amplitudes in the multi-Regge kinematics are equal to the FKL amplitudes
\cite{ptlip}.

\section{Next-to-leading corrections in the forward-rapidity region}
\label{sec:tre}

\begin{figure}[hbt]
\vspace*{-10.cm}
\hspace*{-1.cm}
\epsfxsize=15cm \epsfbox{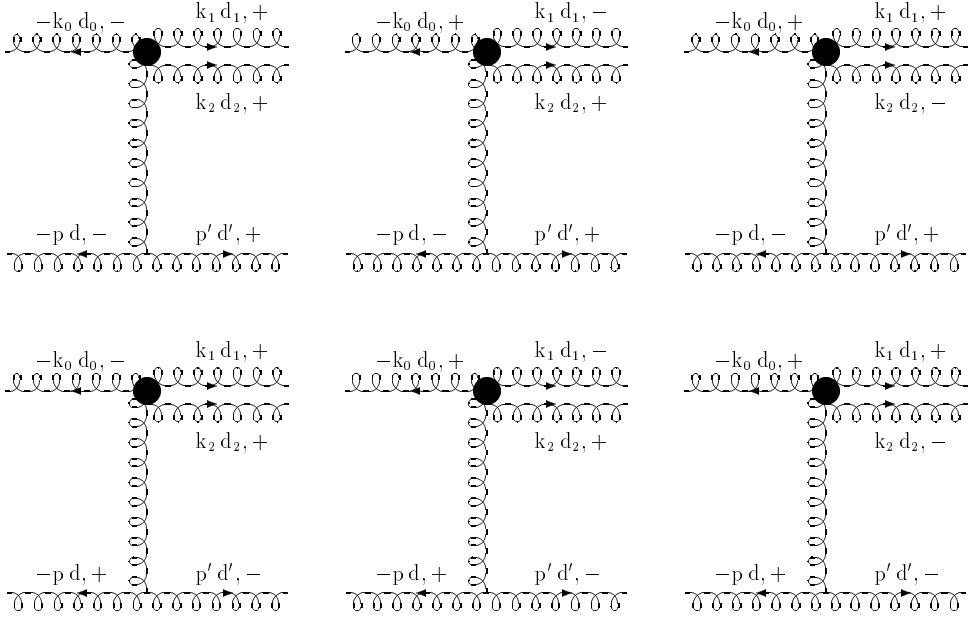}
\vspace*{-3.cm}
\caption{Leading helicity configurations of the 
3-gluon production amplitude, with 2 negative-helicity gluons. The gluons are
labelled by their momenta, always taken as outgoing, their colors and
helicities. Gluons $k_1$ and $k_2$ are produced 
in the forward-rapidity region of gluon $k_0$.}
\label{fig:hel}
\end{figure}

In order to compute the next-to-leading corrections to the FKL amplitudes
in the forward-rapidity region, we consider the production of 3 gluons of 
momenta $k_1$, $k_2$ and $p'$ in the scattering between two gluons of 
momenta $k_0$ and $p$, with gluons $k_1$ and $k_2$ in the forward-rapidity
region of gluon $k_0$ (Fig. \ref{fig:hel}); thus we require that gluons 
$k_1$ and $k_2$ have likewise rapidity, but much larger than the one of $p'$, 
and that they all have comparable transverse momenta,
\begin{equation}
y_1 \simeq y_2 \gg y'\,;\qquad |k_{1\perp}|\simeq|k_{2\perp}|\simeq|p'_{\perp}|
\, .\label{qmreg}
\end{equation}
We are going to show that the amplitude for the production of 3 gluons
in the helicity configuration 
$(-k_0,-\nu_0; k_1,\nu_1; k_2,\nu_2; p',\nu'; -p,-\nu)$ may be written as,
\begin{eqnarray}
& & i\, M(-k_0,-\nu_0; k_1,\nu_1; k_2,\nu_2; p',\nu'; -p,-\nu) \nonumber\\
& & = 2\sqrt{2}\, i\, g^3\, {\hat s\over |p'_{\perp}|^2}\, 
C_{-\nu\nu'}(-p,p')\, \left\{ A_{-\nu_0\nu_1\nu_2}(-k_0,k_1,k_2)
\right. \label{trepos}\\ & & \times
{\rm tr} \left( \lambda^{d_0} \lambda^{d_1} \lambda^{d_2} \lambda^{d'} 
\lambda^d - \lambda^{d_0} \lambda^{d_1} \lambda^{d_2} \lambda^d \lambda^{d'} + 
\lambda^{d_0} \lambda^{d'} \lambda^d \lambda^{d_2} \lambda^{d_1} - 
\lambda^{d_0} \lambda^d \lambda^{d'} \lambda^{d_2} \lambda^{d_1} \right) 
\nonumber\\ & &
\left. - B_{-\nu_0\nu_1\nu_2}(-k_0,k_1,k_2)\, {\rm tr} \left(
\lambda^{d_0} \lambda^{d_1} \lambda^{d'} \lambda^d \lambda^{d_2}
- \lambda^{d_0} \lambda^{d_2} \lambda^d \lambda^{d'} \lambda^{d_1} \right) 
+ (1 \leftrightarrow 2) \right\}\, ,\nonumber
\end{eqnarray}
with the production vertex $C$ of gluon $p'$ determined by
the first of eq.(\ref{contra}),
\begin{equation}
C_{-+}(-p,p') = {{p'}_{\perp}^* \over p'_{\perp}} \qquad
C_{+-}(-p,p') = C_{-+}^*(-p,p')\, ,\label{treposb}
\end{equation}
and with the vertices $A$ and $B$ computed in sect.~\ref{sec:treone}. 
The following considerations motivate the analytic form of eq.(\ref{trepos}):
because of the large rapidity interval between gluons $k_1$
and $k_2$ and gluon $p'$ we expect the amplitude to be dominated by gluon
exchange in the crossed channel, and thus to scale like $\hat s/
|{p'}_{\perp}|^2$, $-|{p'}_{\perp}|^2$ being the momentum transfer $\hat t$
(\ref{frinv}) (Appendix \ref{sec:appc}). In addition, we write 
eq.(\ref{trepos}) in such a way to stress that the vertex $C$
of gluon $p'$ and the vertices $A$ and $B$ of gluons $k_1$ 
and $k_2$, which the amplitude factorizes into, transform separately
into their complex conjugates under the respective helicity reversal,
as the whole amplitude does under the overall helicity reversal.

\subsection{The Parke-Taylor amplitudes}
\label{sec:treone}

\begin{figure}[hbt]
\vspace*{0cm}
\centerline{\epsfxsize=6cm \epsfbox{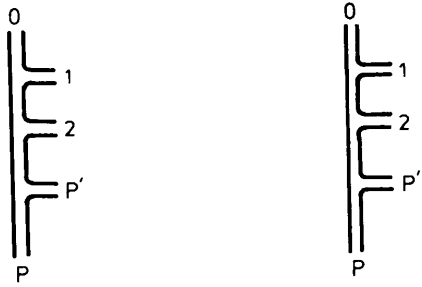}}
\vspace*{0.cm}
\caption{3-gluon production amplitude in the color ordering ($a$)
[$0,1,2,p',p$] and ($b$) ($1\leftrightarrow 2$).}
\label{fig:one}
\end{figure}

We fix $\tilde{p}_A = -k_0$ and $\tilde{p}_B = -p$ in eq.(\ref{one}). From 
the spinor products
(\ref{frpro}) (Appendix \ref{sec:appc}) and eq.(\ref{two}) we note that the
leading helicity configurations are the ones for which the pair of
negative-helicity gluons is one of the following (Fig.~\ref{fig:hel}),
\begin{equation}
(-p,-k_0),\quad (-p,k_i),\quad (p',-k_0),\quad (p',k_i), \label{neg}
\end{equation}
with $i=1,2$, and to start out we consider the pair $(-p,-k_0)$
(Fig. \ref{fig:hel}a). From eq.(\ref{two}) we have,
\begin{equation}
i\, m(-k_0,-; k_1,+; k_2,+; p',+; -p,-) = 4\sqrt{2}\, i\, g^3\, 
{\langle k_0 p\rangle^4\over \langle k_0 k_1\rangle \langle k_1 k_2\rangle
\langle k_2 p'\rangle \langle p' p\rangle \langle p k_0\rangle}\, .\label{twob}
\end{equation}
We insert eq.(\ref{twob}) back into eq.(\ref{one}) and examine 
all the color orderings, as in ref.~\cite{ptlip} and \cite{vd}.
We start with the ordering [$0,1,2,p',p$] (Fig.\ref{fig:one}a).
Using the spinor products (\ref{frpro}) (Appendix \ref{sec:appc}), the 
identity (\ref{flip}) (Appendix \ref{sec:appa}) and eq.(\ref{treposb}),
we obtain
\begin{eqnarray} 
{\rm coeff.\; of\; tr}\left(\lambda^{d_0} \lambda^{d_1} \lambda^{d_2}
\lambda^{d'} \lambda^d\right) &\equiv& 2\sqrt{2}\, i\, g^3\, {\hat s 
\over |p'_{\perp}|^2}\, C_{-+}(-p,p')\, A_{-++}(-k_0,k_1,k_2) \label{tre}\\
A_{-++}(-k_0,k_1,k_2) &=& 2\, {p'_{\perp}\over k_{1\perp}}
{1\over k_{2\perp} - k_{1\perp} {k_2^+\over k_1^+}}\, ,\nonumber
\end{eqnarray}
which has a massless divergence when gluons $1$ and $2$ become collinear.
In the multi-Regge limit, $y_1\gg y_2$, the second term in the denominator
of $A_{-++}$ drops out, and eq.(\ref{tre}) is in agreement with the 
corresponding color ordering of eq.(\ref{fklh}), for $n=1$.
We keep then fixed the position of gluons $k_0$ and $p$ 
in the color ordering and permute the outgoing gluons. The contribution of 
the ordering [$0,2,1,p',p$] (Fig.\ref{fig:one}b) is obtained by exchanging 
the labels
$1$ and $2$ in eq.(\ref{tre}), however the ensuing coefficient is subleading
in the multi-Regge limit, $y_1\gg y_2$. Every other color configuration 
with all the gluons on the front of the lego plot is 
subleading to the required accuracy. 

\begin{figure}[hbt]
\vspace*{0cm}
\centerline{\epsfxsize=10cm \epsfbox{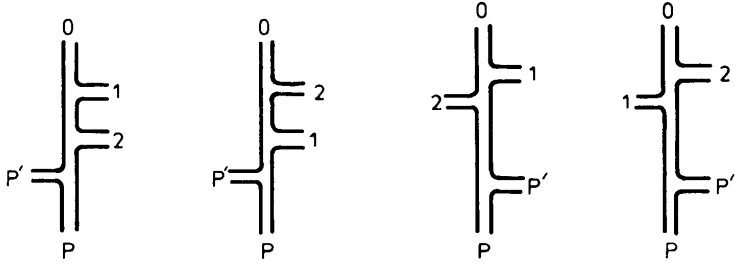}}
\vspace*{0.cm}
\caption{3-gluon production amplitude in the color ordering ($a$)
[$0,1,2,p,p'$] and ($b$) ($1\leftrightarrow 2$); ($c$) [$0,1,p',p,2$]
and ($d$) ($1\leftrightarrow 2$).}
\label{fig:due}
\end{figure}

Next, we move gluon $p$ one step to the left and consider the color orderings 
[$0,1,2,p,p'$] and ($1\leftrightarrow 2$). These correspond to untwisting 
the color lines as done in Fig.~\ref{fig:double}, and having gluon $p'$ on 
the back of the two-sided plot (Fig.\ref{fig:due}a and \ref{fig:due}b). 
Using the spinor products (\ref{frpro}) (Appendix \ref{sec:appc}) we find,
\begin{equation}
\langle k_0 k_1\rangle \langle k_1 k_2\rangle
\langle k_2 p\rangle \langle p p'\rangle \langle p' k_0\rangle = -
\langle k_0 k_1\rangle \langle k_1 k_2\rangle
\langle k_2 p'\rangle \langle p' p\rangle \langle p k_0\rangle\, ,\label{trec}
\end{equation}
with the product on the right-hand side computed in eq.(\ref{tre}).
Then we take gluon $2$
to the back of the two-sided plot (Fig.\ref{fig:due}c) and consider the color
ordering [$0,1,p',p,2$]. The spinor products yield,
\begin{eqnarray}
{\rm coeff.\; of\; tr}\left(\lambda^{d_0} \lambda^{d_1} \lambda^{d'} \lambda^d 
\lambda^{d_2}\right) &\equiv& - 2\sqrt{2}\, i\, g^3\, {\hat s \over 
|p'_{\perp}|^2}\, C_{-+}(-p,p')\, B_{-++}(-k_0,k_1,k_2) \label{tred}\\ 
B_{-++}(-k_0,k_1,k_2) 
&=& 2\, {p'_{\perp}\over k_{1\perp} k_{2\perp}}\, .\nonumber
\end{eqnarray}
The color ordering of eq.(\ref{tred}) does not contain the massless 
divergence when gluons $1$ and $2$ become collinear and
is in agreement with the corresponding
color ordering in the multi-Regge limit (\ref{fklh}), for $n=1$. 
The fact that the color ordering [$0,1,p',p,2$] is
finite in the collinear limit $2k_1\cdot k_2\rightarrow 0$,
whether we compute it in the multi-Regge kinematics (\ref{mreg}) or in the
next-to-leading corrections to it (\ref{qmreg}), means that, in order
to exhibit a collinear divergence, gluons $1$ and $2$ must be produced with 
equal rapidity and azimuthal angle {\it and} be adjacent in color space,
or, in lego-plot language, that the two gluons must
be overlapping on the plot {\it and} be produced on the same side.
Finally, eq.(\ref{tred}) is invariant under exchange of the labels $1$ and 
$2$, thus the color ordering [$0,2,p',p,1$] (Fig.\ref{fig:due}d)
yields the same contribution as in eq.(\ref{tred}), and in particular,
\begin{equation}
B_{-++}(-k_0,k_1,k_2) = A_{-++}(-k_0,k_1,k_2) + A_{-++}(-k_0,k_2,k_1)\,
.\label{baa}
\end{equation}
The remaining two color configurations with one
gluon on the back of the two-sided plot are subleading.

The other color configurations, with two or three gluons on the back
of the two-sided plot, are obtained by taking the color orderings 
[$0,1,2,p',p$], [$0,1,2,p,p'$], [$0,1,p',p,2$], and ($1\leftrightarrow 2$)
in reverse order. Because of the ciclicity of the traces and
the identity (\ref{flip}) (Appendix \ref{sec:appa}), this yields the
same result as in eq.(\ref{tre}), (\ref{trec}) and (\ref{tred}) but
with opposite sign. 

Substituting then eq.(\ref{tre}-\ref{tred}) into
eq.(\ref{one}), we obtain the 3-gluon production amplitude in the 
helicity configuration (\ref{twob}) in the form of eq.(\ref{trepos})
with $C_{-+}$, $A_{-++}$ and $B_{-++}$ given in eq.(\ref{treposb}),
(\ref{tre}) and (\ref{tred}) respectively.
In the multi-Regge limit, $y_1\gg y_2$, the 4 color configurations obtained
by exchanging gluons $k_1$ and $k_2$ on the same side of the lego plot
(Fig.\ref{fig:one}b and \ref{fig:due}b and analogous ones with front and 
back of the lego plot exchanged) become subleading. Thus out of the 12 color 
configurations of eq.(\ref{trepos}), only the 8 configurations given by 
eq.(\ref{fpro}) for $n=1$ survive, and
eq.(\ref{trepos}) is reduced to eq.(\ref{fklh}).

Then we consider the other helicity configurations (Fig.~\ref{fig:hel}b-f) in
eq.(\ref{neg}). Substituting the product $\langle k_0 p\rangle^4$ in
eq.(\ref{twob}) with the suitable product according to the pair of
negative-helicity gluons considered, we obtain,
\begin{eqnarray}
& & M(-k_0,+; k_1,-; k_2,+; p',+; -p,-) \nonumber\\ & & = {1\over\left(1+
{k_2^+\over k_1^+} \right)^2}\,
M(-k_0,-; k_1,+; k_2,+; p',+; -p,-) \label{foura}\\
& & M(-k_0,-; k_1,+; k_2,+; p',-; -p,+) \nonumber\\ & & = \left({p'_{\perp}
\over {p'}_{\perp}^*}\right)^2 M(-k_0,-;k_1,+;k_2,+;p',+;-p,-) \label{fourb}\\
& & M(-k_0,+; k_1,-; k_2,+; p',-; -p,+) \nonumber\\ & & = \left({p'_{\perp}
\over {p'}_{\perp}^*}\right)^2 M(-k_0,+;k_1,-;k_2,+;p',+;-p,-)\, .\label{fourc}
\end{eqnarray}
In the multi-Regge limit, $y_1\gg y_2$, eq.(\ref{foura}) (Fig.~\ref{fig:hel}b)
agrees with eq.(\ref{fklhel}). Eq.(\ref{fourb}) (Fig.~\ref{fig:hel}d)
is already in agreement with its multi-Regge limit,
eq.(\ref{fklhelb}), since the amplitude factorizes and the
lower vertex is insensitive to the next-to-leading corrections in the
upper vertex. Eq.(\ref{fourc}) 
(Fig.~\ref{fig:hel}e) is obtained combining the results of eq.(\ref{foura})
and eq.(\ref{fourb}), and in the multi-Regge limit is reduced to 
eq.(\ref{fklhelb}). The configurations $(-k_0,+; k_1,+; k_2,-; p',+; 
-p,-)$ (Fig.~\ref{fig:hel}c) and $(-k_0,+; k_1,+; k_2,-; p',-; -p,+)$
(Fig.~\ref{fig:hel}f) are obtained by exchanging the labels $1$ and $2$ in 
eq.(\ref{foura}) and (\ref{fourc}) respectively; however, in the multi-Regge 
limit $y_1\gg y_2$, they are subleading since then helicity is not conserved
in the production vertex of gluon $k_1$. Thus, 4 out of the 6 helicity 
configurations of eq.(\ref{neg}) survive in the multi-Regge limit, in
agreement with ref.~\cite{ptlip}. 

\begin{figure}[hbt]
\vspace*{-7.5cm}
\hspace*{0.cm}
\epsfxsize=15cm \epsfbox{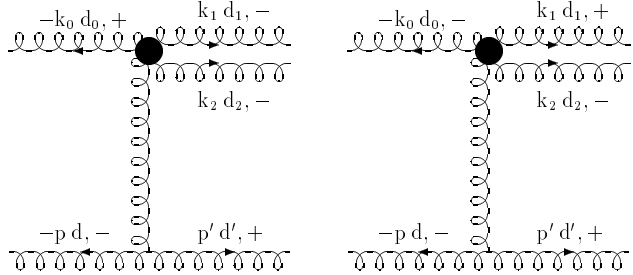}
\vspace*{-10.cm}
\caption{3-gluon production amplitude, with 2 positive-helicity gluons.}
\label{fig:helpos}
\end{figure}

The configurations with only two positive-helicity gluons, which correspond 
to inverting the helicity of all the gluons in Fig.~\ref{fig:hel}, are obtained
from the ones of eq.(\ref{neg}) by replacing the $\langle p k\rangle$
products with $\left[k p\right]$ products, and by using
eq.(\ref{flips}) (Appendix \ref{sec:appa}).
Thus they may be computed by taking the complex conjugate of the spinor 
products (\ref{tre}), (\ref{trec}) and (\ref{tred}), and by using 
eq.(\ref{foura}), (\ref{fourb}) and (\ref{fourc}). Added to the 6
helicity configurations (\ref{neg}), they cover all the 12 leading
helicity configurations of the 3-gluon amplitude in the kinematics 
(\ref{qmreg}). In particular we consider two examples: 
the configuration $(-k_0,+; k_1,-; k_2,-; p',+; -p,-)$ 
(Fig.~\ref{fig:helpos}a), which corresponds to reversing all the helicities in 
the production vertex of gluons $k_1$ and $k_2$ (Fig.~\ref{fig:hel}a),
and is obtained by taking the complex conjugate of the spinor products 
in eq.(\ref{fourb}). The amplitude then assumes the form of 
eq.(\ref{trepos}), with $A_{+--}(-k_0,k_1,k_2) = A_{-++}^*(-k_0,k_1,k_2)$
and $B_{+--}(-k_0,k_1,k_2) = B_{-++}^*(-k_0,k_1,k_2)$, in agreement with
the discussion following eq.(\ref{trepos}); the configuration 
$(-k_0,-; k_1,+; k_2,-; p',+; -p,-)$ (Fig.~\ref{fig:helpos}b), which 
corresponds to reversing the helicity of one 
of the outgoing gluons in the upper vertex (Fig.~\ref{fig:hel}a), and 
is obtained by taking the complex conjugate of the spinor products in
eq.(\ref{fourc}). The amplitude
has the form of eq.(\ref{trepos}), with $A_{-+-}(-k_0,k_1,k_2) = 
A_{+-+}^*(-k_0,k_1,k_2)$ and $B_{-+-}(-k_0,k_1,k_2) = B_{+-+}^*(-k_0,k_1,k_2)$ 
and $A_{+-+}(-k_0,k_1,k_2)$ and $B_{+-+}(-k_0,k_1,k_2)$ defined by
eq.(\ref{foura}). By taking then the multi-Regge limit of eq.(\ref{trepos})
with $(\nu_0\nu_1\nu_2) = (-+-)$ and $(\nu_0\nu_1\nu_2) = (-++)$ we obtain,
\begin{eqnarray}
& & \lim_{y_1\gg y_2} M(-k_0,-; k_1,+; k_2,-; p',+; -p,-) \label{frchh}\\
& & = {{p'}_{\perp}^* k_{1\perp} k_{2\perp}\over p'_{\perp} k^*_{1\perp} 
k^*_{2\perp}}\, \lim_{y_1\gg y_2} M(-k_0,-; k_1,+; k_2,+; p',+; -p,-)\, 
,\nonumber
\end{eqnarray} 
in agreement with the helicity flip in the Lipatov vertex of gluon $k_2$
in the FKL amplitudes, eq.(\ref{chh}), with $q_{1\perp}\equiv -k_{1\perp}$,
$q_{2\perp} \equiv p'_{\perp}$ and $p_{1\perp}\equiv k_{2\perp}$.

The extension of the kinematics (\ref{qmreg}) to the production of
$n+2$ gluons, with 2 gluons with likewise rapidity in the forward-rapidity 
region of the upper vertex (\ref{qmrapp}-\ref{frpapp}) 
(Appendix \ref{sec:appc}), is straightforward. 
Generalizing equations (\ref{tre}-\ref{tred}) to the production of $n+2$ 
gluons and using them to compute the other color configurations, the amplitude
for the extension of the helicity configuration of Fig.~\ref{fig:hel}a is,
\begin{eqnarray}
& & i\, M(-k_0,-; k_1,+; k_2,+; p_1,+;...; p_n,+; -p,-) \nonumber\\ & & =
(-1)^{n+1}\, 2^{2+n/2}\, i\, g^{n+2}\, \hat s\,
{1\over \prod_{i=1}^n p_{i\perp}} 
\left\{ \left[ {1\over k_{1\perp}} {1\over k_{2\perp} - k_{1\perp} {k_2^+
\over k_1^+}} \right.\right. \label{fourn}\\ & & \times
{\rm tr} \left( \lambda^{d_0} \lambda^{d_1} \lambda^{d_2} \lambda^{p_1} \cdots 
\lambda^{p_n} \lambda^p - \sum_{i=1}^n \lambda^{d_0} \lambda^{d_1} 
\lambda^{d_2} \lambda^{p_1} \cdots \lambda^{p_{i-1}} \lambda^{p_{i+1}}
\cdots \lambda^{p_n} \lambda^p \lambda^{p_i} \right. \nonumber\\ & & \left.
+ \lambda^{d_0} \lambda^{p_1} \cdots \lambda^{p_n} \lambda^p \lambda^{d_2} 
\lambda^{d_1} + \sum_{i<j} \lambda^{d_0} \lambda^{d_1} \lambda^{d_2} 
\lambda^{p_1} \cdots \lambda^{p_{i-1}} \lambda^{p_{i+1}}\cdots\lambda^{p_{j-1}}
\lambda^{p_{j+1}} \cdots \lambda^{p_n} \lambda^p \lambda^{p_j} \lambda^{p_i} + 
... \right) \nonumber\\ & & \left. - {1\over k_{1\perp} k_{2\perp}} {\rm tr} 
\left( \lambda^{d_0} \lambda^{d_1} \lambda^{p_1} \cdots \lambda^{p_n} \lambda^p
\lambda^{d_2} - \sum_{i=1}^n \lambda^{d_0} \lambda^{d_1} \lambda^{p_1} \cdots 
\lambda^{p_{i-1}} \lambda^{p_{i+1}}\cdots \lambda^{p_n} \lambda^p \lambda^{p_i}
\lambda^{d_2} + ... \right) \right] \nonumber\\ & & \left.
+ (1 \leftrightarrow 2)\, \right\}\, ,\nonumber
\end{eqnarray}
where we have explicitly shown the color configurations with $n+2$, $n+1$ and
$n$ gluons on the front of the lego plot.
By explicit counting, we see that the leading
color configurations in eq.(\ref{fourn}) are $3!\, 2^n$, which for $n=1$
yields the 12 color configurations of eq.(\ref{trepos}). The other helicity
configurations with 2 negative or 2 positive-helicity gluons
may be accordingly generalized. 

\subsection{The Fadin-Lipatov amplitudes}
\label{sec:tredue}

The 3-gluon production amplitude, with gluons $k_1$ and $k_2$ produced
in the forward-rapidity region of gluon $k_0$, as specified by the kinematics 
(\ref{qmreg}), has been computed in ref.~\cite{fl}, and for a generic
helicity configuration is given by,
\begin{eqnarray}
& & i\, M^{d_0d_1d_2d'd}_{\nu_0\nu_1\nu_2\nu'\nu} =  
\epsilon_{\mu_0}^{\nu_0*}(k_0) \epsilon_{\mu_1}^{\nu_1}(k_1)
\epsilon_{\mu_2}^{\nu_2}(k_2) \epsilon_{\mu}^{\nu*}(p) 
\epsilon_{\mu'}^{\nu'}(p')\, 4\, g^3\, {1\over\hat t} f^{d'cd}
\Gamma^{\mu\mu'}(p,p',-k_0) \nonumber\\ & & \times \left\{ 
\Gamma^{\mu_0\mu_1}(-k_0,k_1,p) \left[ f^{d_0c'd_2} f^{d_1c'c} 
D^{\mu_2}(-k_0,k_1,k_2) + f^{d_0c'c} f^{d_1c'd_2} D^{\mu_2}(k_1,-k_0,k_2) 
\right] \right. \nonumber\\ & & \left. + \left(\begin{array}{c} 
k_1\leftrightarrow k_2\\ \nu_1\leftrightarrow \nu_2\\ d_1\leftrightarrow d_2 
\end{array}\right) + \left(\begin{array}{c} -k_0\leftrightarrow k_2\\ 
\nu_0\leftrightarrow \nu_2\\ d_0\leftrightarrow d_2 \end{array}\right) \right\}
\, ,\label{five}
\end{eqnarray}
where the $\Gamma$-tensors are given in eq.(\ref{gamm}), and the 2-gluon
production vertex $D$ in the forward-rapidity region is \cite{fl},
\begin{eqnarray}
& & D^{\mu}(k_0,k_1,k_2) \label{six}\\ & & = {1\over k_0\cdot k_1} \left[ 
\left(k_1\cdot k_2 - p\cdot p' {k_1\cdot p\over k_2\cdot p} \right) p^{\mu} + 
{k_1\cdot p\over k_0\cdot k_2} (k_1\cdot k_2 + p\cdot p')\, k_0^{\mu} + 
k_0\cdot p\, k_1^{\mu}\right]\, ,\nonumber
\end{eqnarray}
with the Mandelstam invariants given in eq.(\ref{frinv}) (Appendix 
\ref{sec:appc}). The $D$-vertex is gauge invariant with respect to the 
last of its arguments, 
\begin{equation}
D(-k_0,k_1,k_2)\cdot k_2 =0\, ,\label{gauged}
\end{equation}
where we have used the $+$ momentum conservation (\ref{frkin}) (Appendix 
\ref{sec:appc}). Using eq.(\ref{gauged}) and the gauge invariance of the
$\Gamma$-tensors (\ref{gauge}), one can see that the amplitude (\ref{five}) 
is invariant with respect to arbitrary gauge transformations \cite{fl}, 
\cite{lipatov}. The symmetry
of the amplitude (\ref{five}) under permutations of the gluons in the
forward-rapidity region is manifest.

In order to make the color structure explicit, we use eq.(\ref{fpro}) and
rewrite eq.(\ref{five}) as
\begin{eqnarray}
& & i\, M^{d_0d_1d_2d'd}_{\nu_0\nu_1\nu_2\nu'\nu} = - 
\epsilon_{\mu_0}^{\nu_0*}(k_0) \epsilon_{\mu_1}^{\nu_1}(k_1)
\epsilon_{\mu_2}^{\nu_2}(k_2) \epsilon_{\mu}^{\nu*}(p) 
\epsilon_{\mu'}^{\nu'}(p')\, 8\,(-i)^3\, g^3\, {1\over\hat t}
\Gamma^{\mu\mu'}(p,p',-k_0) \nonumber\\ & & \times \left\{ \left[
\Gamma^{\mu_0\mu_1}(-k_0,k_1,p) D^{\mu_2}(-k_0,k_1,k_2) -
\Gamma^{\mu_2\mu_1}(k_2,k_1,p) D^{\mu_0}(k_2,k_1,-k_0) \right] 
\right. \nonumber\\ & & \times{\rm tr}\left(\lambda^{d_0}\,\left[\lambda^{d_2},
\left[\lambda^{d_1},\left[\lambda^{d'},\lambda^d\right]\right]\right]\right) 
\label{seven} \\ & & +
\left[ \Gamma^{\mu_0\mu_1}(-k_0,k_1,p) D^{\mu_2}(k_1,-k_0,k_2) -
\Gamma^{\mu_0\mu_2}(-k_0,k_2,p) D^{\mu_1}(k_2,-k_0,k_1) \right] \nonumber\\ & &
\times {\rm tr}\left(\lambda^{d_1}\,\left[\lambda^{d_2},\left[\lambda^{d_0},
\left[\lambda^{d'},\lambda^d\right]\right]\right]\right) \nonumber\\ & & +
\left[ \Gamma^{\mu_0\mu_2}(-k_0,k_2,p) D^{\mu_1}(-k_0,k_2,k_1) -
\Gamma^{\mu_2\mu_1}(k_2,k_1,p) D^{\mu_0}(k_1,k_2,-k_0) \right] \nonumber\\ & &
\left. \times {\rm tr}\left(\lambda^{d_0}\,\left[\lambda^{d_1},\left[
\lambda^{d_2},\left[\lambda^{d'},\lambda^d\right]\right]\right]\right) \right\}
\, .\nonumber
\end{eqnarray}
Eq.(\ref{seven}) simplifies considerably once
the gluon helicities are fixed. For sake of comparison with the PT amplitudes,
we consider the configuration 
with helicities $\nu_0 = \nu_1 = \nu_2 = +$. We perform the contractions of the
helicity-conserving tensors (\ref{gamm}) with the gluon polarizations
using eq.(\ref{frexh}) (Appendix \ref{sec:appc}). For the incoming gluons 
they of course coincide with eq.(\ref{contra}), 
\begin{eqnarray}
\Gamma^{\mu\mu'}(p,p',-k_0)\, \epsilon^{+*}_{\mu}(p, k_0)\, \epsilon^+_{\mu'}
(p', k_0) &=& -{{p'}_{\perp}^*\over p'_{\perp}} \nonumber\\
\Gamma^{\mu_0\mu_i}(-k_0,k_i,p)\, \epsilon^{+*}_{\mu_0}(k_0, p)\, 
\epsilon^+_{\mu_i}(k_i, p) &=& -1 \label{frcont}\\ 
\Gamma^{\mu_2\mu_1}(k_2,k_1,p)\, \epsilon^+_{\mu_2}(k_2,p)\, 
\epsilon^+_{\mu_1}(k_1,p) &=& 0\, ,\nonumber
\end{eqnarray}
with $i=1,2$. Accordingly, eq.(\ref{seven}) becomes
\begin{eqnarray}
& & i\, M(k_0,+; k_1,+; k_2,+; p',\nu'; p,\nu) = - 8\,(-i)^3\, g^3\, 
{1\over\hat t}\, C_{\nu\nu'}(p,p') \nonumber\\ & & \times \left\{
D(-k_0,k_1,k_2)\cdot \epsilon^+(k_2,p)\, 
{\rm tr}\left(\lambda^{d_0}\,\left[\lambda^{d_2},\left[\lambda^{d_1},
\left[\lambda^{d'},\lambda^d\right]\right]\right]\right) \right. \label{otto}\\
& & + \left[ D(k_1,-k_0,k_2)\cdot \epsilon^+(k_2,p)\, -
D(k_2,-k_0,k_1)\cdot \epsilon^+(k_1,p) \right] 
{\rm tr}\left(\lambda^{d_1}\,\left[\lambda^{d_2},\left[\lambda^{d_0},
\left[\lambda^{d'},\lambda^d\right]\right]\right]\right) \nonumber\\
& & \left. + D(-k_0,k_2,k_1)\cdot \epsilon^+(k_1,p)
{\rm tr}\left(\lambda^{d_0}\,\left[\lambda^{d_1},\left[\lambda^{d_2},
\left[\lambda^{d'},\lambda^d\right]\right]\right]\right) \right\}\, ,\nonumber
\end{eqnarray}
with the $C$-vertex defined like in eq.(\ref{treposb}),
\begin{equation}
C_{++}(p,p') = {{p'}_{\perp}^* \over p'_{\perp}} \qquad
C_{--}(p,p') = C_{++}^*(p,p')\, .\label{ottob}
\end{equation}
Using eq.(\ref{fpro}), we unfold the nested commutators in eq.(\ref{otto}),
where it is convenient to rewrite the second as, 
\begin{equation}
{\rm tr}\left(\lambda^{d_1}\,\left[\lambda^{d_2},\left[
\lambda^{d_0},\left[\lambda^{d'},\lambda^d\right]\right]\right]\right) =
{\rm tr}\left(\lambda^{d_0}\,\left[ \left[\lambda^{d_2},\lambda^{d_1}\right],
\left[\lambda^{d'},\lambda^d\right] \right]\right)\, .\label{quattor}
\end{equation}
The color orderings we obtain are the same as in eq.(\ref{trepos}), and
we compute their coefficients starting with the one of
${\rm tr}\left(\lambda^{d_0}\lambda^{d_1}\lambda^{d_2}\lambda^{d'}
\lambda^d\right)$. Performing the contractions of the $D$-vertices (\ref{six})
with the gluon polarizations (\ref{frexh}) (Appendix \ref{sec:appc}),
we obtain, after some algebraic manipulations,
\begin{eqnarray}
& & {\rm coeff.\; of}\; {\rm tr}\left(\lambda^{d_0}\lambda^{d_1}
\lambda^{d_2}\lambda^{d'} \lambda^d\right) =  - 8\,(-i)^3\, g^3\, 
{1\over\hat t}\, C_{\nu\nu'}(p,p') \nonumber\\ & & \times
\left[D(-k_0,k_2,k_1)\cdot \epsilon^+(k_1,p) - D(k_1,-k_0,k_2)\cdot 
\epsilon^+(k_2,p)\, + D(k_2,-k_0,k_1)\cdot \epsilon^+(k_1,p) \right]
\nonumber\\ & & = 4\sqrt{2}i\, g^3\, {\hat s\over |p'_{\perp}|^2}
C_{\nu\nu'}(p,p')\, {p'_{\perp}\over k_{1\perp}}
{1\over k_{2\perp} - k_{1\perp} {k_2^+\over k_1^+}}\, .\label{nove}
\end{eqnarray}
In addition, eq.(\ref{otto}) is manifestly antisymmetric under the
exchange $d\leftrightarrow d'$, thus
\begin{equation}
{\rm coeff.\; of}\; {\rm tr}\left(\lambda^{d_0}\lambda^{d_1}
\lambda^{d_2}\lambda^d \lambda^{d'}\right) = -
{\rm coeff.\; of}\; {\rm tr}\left(\lambda^{d_0}\lambda^{d_1}
\lambda^{d_2}\lambda^{d'} \lambda^d\right)\, ,\label{ten}
\end{equation}
plus contributions with the labels of gluons $1$ and $2$ exchanged in
eq.(\ref{nove}) and (\ref{ten}).
The coefficient of ${\rm tr}\left(\lambda^{d_0}\lambda^{d_1}
\lambda^{d'} \lambda^d \lambda^{d_2}\right)$ is symmetric under the
exchange of gluons $1$ and $2$,
\begin{eqnarray}
& & {\rm coeff.\; of}\; {\rm tr}\left(\lambda^{d_0}\lambda^{d_1}
\lambda^{d'} \lambda^d \lambda^{d_2}\right) \nonumber\\ & & =
8\,(-i)^3\, g^3\, {1\over\hat t}\, C_{\nu\nu'}(p,p')\, 
\left[ D(-k_0,k_1,k_2)\cdot \epsilon^+(k_2,p) +
D(-k_0,k_2,k_1)\cdot \epsilon^+(k_1,p) \right] \nonumber\\ & &
= - 4\sqrt{2}\, i\, g^3\, {\hat s\over |p'_{\perp}|^2}
C_{\nu\nu'}(p,p')\, {p'_{\perp}\over k_{1\perp} k_{2\perp}}\, .\label{elev}
\end{eqnarray}
The contraction,
\begin{equation}
D(-k_0,k_1,k_2)\cdot \epsilon^+(k_2,p) = - {k_0\cdot p\over\sqrt{2}} 
\left({k^*_{1\perp}\over k_0\cdot k_1} + {k^*_{2\perp}\over k_0\cdot k_2} 
{k_1^+\over k_2^+} \right)\, \label{twel}
\end{equation}
used in eq.(\ref{elev}), is proportional in the multi-Regge limit to the
Lipatov vertex,
\begin{eqnarray}
\lim_{y_1\gg y_2} D(-k_0,k_1,k_2)\cdot \epsilon^+(k_2,p) &=& {1\over\sqrt{2}}
{k_0\cdot p\over k_0\cdot k_1} {k^*_{1\perp} p'_{\perp} \over k_{2\perp}} 
\label{tredic}\\ &=& - {1\over 2} {k_0\cdot p\over k_0\cdot k_1}\,
\epsilon^+(k_2, p)\cdot C(q_1,q_2)\, ,\nonumber
\end{eqnarray}
in agreement with ref.~\cite{fl}. In eq.(\ref{tredic})
we have used the multi-Regge limit (\ref{mrinv}) (Appendix 
\ref{sec:appb}) of the invariants (\ref{frinv}) (Appendix \ref{sec:appc}), 
and eq.(\ref{verc}) with $q_{1\perp}\equiv -k_{1\perp}$, $q_{2\perp}
\equiv p'_{\perp}$ and $p_{1\perp}\equiv k_{2\perp}$. 

In the two-sided lego-plot picture,
${\rm tr}\left(\lambda^{d_0}\lambda^{d_1}
\lambda^{d_2}\lambda^{d'}\lambda^d\right)$ plus ($1\leftrightarrow 2$),
yields the leading color configurations with all the gluons on the 
front of the lego plot (Fig.\ref{fig:one}); ${\rm tr}\left(\lambda^{d_0}
\lambda^{d_1}\lambda^{d_2}\lambda^d \lambda^{d'}\right)$ and ${\rm tr}\left(
\lambda^{d_0}\lambda^{d_1}\lambda^{d'} \lambda^d \lambda^{d_2}\right)$
plus ($1\leftrightarrow 2$), yield the leading color 
configurations with one gluon on the back of the lego plot (Fig.\ref{fig:due}).
The coefficients of the color configurations with two and three gluons on 
the back of the lego plot are obtained by changing the sign to the ones with
three gluons (\ref{nove}) and two gluons (\ref{ten}), (\ref{elev}) on the 
front of the lego plot, as can be seen by direct inspection of the unfolded
nested commutators in eq.(\ref{otto}). Thus, using eq.(\ref{nove}-\ref{elev}), 
eq.(\ref{otto}) is reduced to eq.(\ref{trepos}), proving the equivalence
of the Fadin-Lipatov and the helicity amplitudes for the configuration
of eq.(\ref{twob}). Analogously all the other leading helicity configurations
may be computed from eq.(\ref{seven}), and compared to the helicity
amplitudes of sect.~\ref{sec:treone}.

\section{Next-to-leading corrections in the central-rapidity region}
\label{sec:four}

In order to compute the next-to-leading corrections to the FKL amplitudes
in the central-rapidity region, we consider the production of 4 gluons of 
momenta $p'_A$, $k_1$, $k_2$ and $p'_B$ in the scattering between two gluons 
of momenta $p_A$ and $p_B$. We require that gluons $k_1$ and $k_2$ have 
likewise rapidity and are separated through large rapidity intervals
from the forward-rapidity regions,
with all the gluons having comparable transverse momenta,
\begin{equation}
y'_A\gg y_1 \simeq y_2 \gg y'_B\,;\qquad |k_{1\perp}| \simeq |k_{2\perp}|
\simeq |p'_{A\perp}| \simeq |p'_{B\perp}|\, .\label{crreg}
\end{equation}
We are going to show that the amplitude for the production of 4 gluons
in the helicity configuration 
$(-p_A,-\nu_A; p'_A,\nu'_A; k_1,\nu_1; k_2,\nu_2; p'_B,\nu'_B; -p_B,-\nu_B)$
has the form,
\begin{eqnarray}
& & i\, M(-p_A,-\nu_A; p'_A,\nu'_A; k_1,\nu_1; k_2,\nu_2; p'_B,\nu'_B; 
-p_B,-\nu_B) \label{centr}\\ & & = - 4\, i\, g^4\, {\hat s 
\over |p'_{A\perp}|^2 |p'_{B\perp}|^2}\, C_{-\nu_A\nu'_A}(-p_A,p'_A)\,
C_{-\nu_B\nu'_B}(-p_B,p'_B) \nonumber\\ & &
\times \left\{ A_{\nu_1\nu_2}(k_1,k_2)
\left[ {\rm tr} \left( \lambda^a \lambda^{a'} \lambda^{d_1} \lambda^{d_2} 
\lambda^{b'} \lambda^b - \lambda^a \lambda^{a'} \lambda^{d_1} 
\lambda^{d_2} \lambda^b \lambda^{b'} \right. \right.\right. \nonumber\\ & & 
\left.\left. - \lambda^a \lambda^{d_1} \lambda^{d_2}
\lambda^{b'} \lambda^b \lambda^{a'} + \lambda^a \lambda^{d_1} \lambda^{d_2}
\lambda^b \lambda^{b'} \lambda^{a'} \right) 
+ {\rm traces}\, {\rm in}\, {\rm reverse}\, {\rm order} \right] 
- B_{\nu_1\nu_2}(k_1,k_2) \nonumber\\ & &
\left. \times \left[ {\rm tr} \left( 
\lambda^a \lambda^{a'} \lambda^{d_1} \lambda^{b'} \lambda^b \lambda^{d_2}
- \lambda^a \lambda^{a'} \lambda^{d_1} \lambda^b \lambda^{b'} \lambda^{d_2} 
\right) + {\rm traces}\, {\rm in}\, {\rm reverse}\, {\rm order} \right]
+ (1 \leftrightarrow 2)_{\nu_1/\nu_2} \right\}\, ,\nonumber
\end{eqnarray}
with the production vertices $C$ of gluons $p'_A$ and $p'_B$ determined by
eq.(\ref{contra}),
\begin{eqnarray}
C_{-+}(-p_A,p'_A) &=& C_{+-}(-p_A,p'_A) = 1 \label{centrc}\\
C_{-+}(-p_B,p'_B) &=& {{p'}_{B\perp}^* \over p'_{B\perp}} \qquad
C_{+-}(-p_B,p'_B) = C_{-+}^*(-p_B,p'_B)\, ,\nonumber
\end{eqnarray}
and with $A_{\nu_1\nu_2}(k_1,k_2)$ and $B_{\nu_1\nu_2}(k_1,k_2)$ computed in
sect.~\ref{sec:fourone} and \ref{sec:fourbis}. In addition,
\begin{equation}
(1 \leftrightarrow 2)_{\nu_1/\nu_2} = \left\{ \begin{array}{ll} 
(1 \leftrightarrow 2) & \mbox{for} \quad \nu_1=\nu_2=\pm \\
(1 \leftrightarrow 2)\, {\cal C} & \mbox{for} \quad \nu_1=-\nu_2=\pm, 
\end{array} \right. \label{exchange}
\end{equation}
with $\cal C$ the complex conjugation. As in eq.(\ref{trepos})
we stress that the two vertices for the production
of gluons $p'_A$ and $p'_B$ in the forward-rapidity regions, and the vertices
of gluons $k_1$ and $k_2$ in the central-rapidity region transform 
separately under the respective helicity reversal. However, differently from 
eq.(\ref{trepos}) we note that in eq.(\ref{centr}) the color structure is 
linked through eq.(\ref{exchange}) to the helicity structure. This is due 
to the fact that besides the PT amplitudes, which describe the production
of gluons $k_1$ and $k_2$ with equal helicities (sect.~\ref{sec:fourone}),
and for which the color and the helicity structures are uncorrelated,
we use, in order to consider the production of gluons $k_1$ and $k_2$ with 
opposite helicities, the amplitudes with 3 negative-helicity gluons
(sect.~\ref{sec:fourbis}), for which the color and the helicity structures 
are intertwined.

\begin{figure}[hbt]
\vspace*{-9.cm}
\hspace*{0.cm}
\epsfxsize=15cm \epsfbox{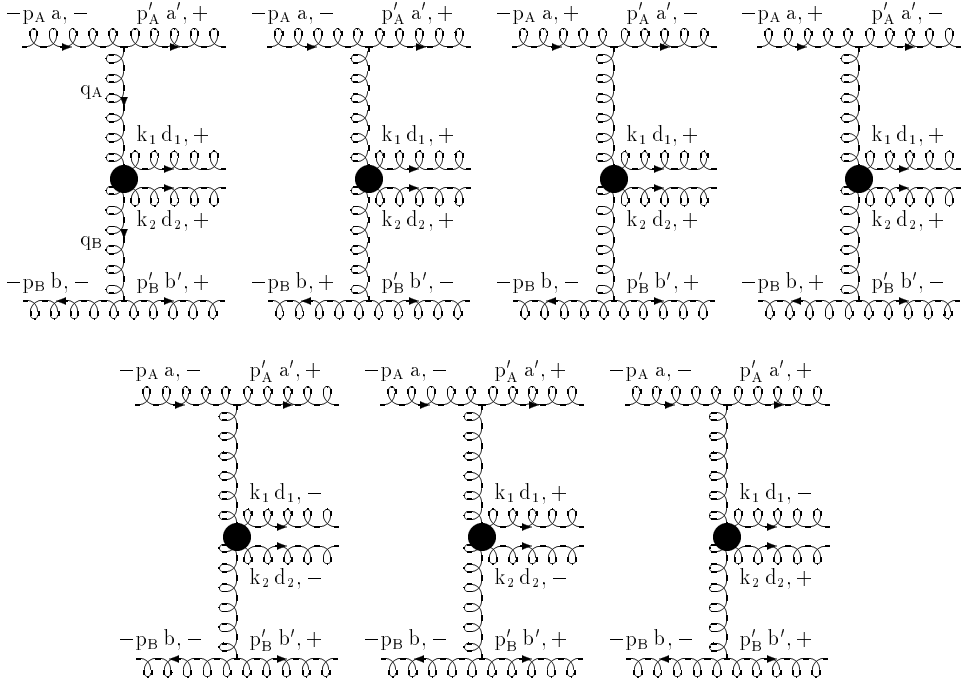}
\vspace*{-3.cm}
\caption{4-gluon production amplitude, with 2 gluons
with likewise rapidity in the central region; ($a-d$) leading helicity 
configurations with 2 negative-helicity gluons; ($e-g$) helicity reversal
in the production vertex of gluons $k_1$ and $k_2$, with respect to the
configuration $a$.} 
\label{fig:cent}
\end{figure}

\subsection{The Parke-Taylor amplitudes}
\label{sec:fourone}

We fix $\tilde{p}_A = -p_A$ and $\tilde{p}_B = -p_B$ in eq.(\ref{one}). From 
the spinor products (\ref{crpro}) (Appendix \ref{sec:appd}) and eq.(\ref{two}),
the leading helicity configurations are the same as in the multi-Regge 
kinematics (Fig.~\ref{fig:cent}a-d) \cite{ptlip}, i.e. the ones for which the 
pair of negative-helicity gluons is one of the following,
\begin{equation}
(-p_A,-p_B),\quad (-p_A,p'_B),\quad (-p_B,p'_A),\quad (p'_A,p'_B)\, 
.\label{crneg}
\end{equation}
As usual, we start with the pair $(-p_A,-p_B)$ (Fig.~\ref{fig:cent}a).
From eq.(\ref{two}) we have,
\begin{eqnarray}
& & i\, m(-p_A,-; p'_A,+; k_1,+; k_2,+; p'_B,+; -p_B,-) \label{cra}\\
& & = 8\, i\, g^4\, {\langle p_A p_B\rangle^4\over \langle p_A p'_A\rangle 
\langle p'_A k_1\rangle \langle k_1 k_2\rangle \langle k_2 p'_B\rangle 
\langle p'_B p_B\rangle \langle p_B p_A\rangle}\, .\nonumber
\end{eqnarray}
\begin{figure}[hbt]
\vspace*{0cm}
\centerline{\epsfxsize=5cm \epsfbox{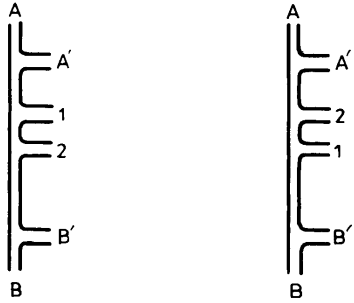}}
\vspace*{0.cm}
\caption{4-gluon production amplitude in the color orderings
[$A,A',1,2,B',B$] and $(1\leftrightarrow 2)$.}
\label{fig:centone}
\end{figure}
We insert eq.(\ref{cra}) into eq.(\ref{one}) and examine 
all the color orderings, starting with\newline [$A,A',1,2,B',B$]
(Fig.~\ref{fig:centone}a) and ($1\leftrightarrow 2$) (Fig.\ref{fig:centone}b).
Using the spinor products (\ref{crpro}) (Appendix \ref{sec:appd}) and
eq.(\ref{centrc}), we obtain
\begin{eqnarray} 
{\rm coeff.\; of}\; {\rm tr}\left(\lambda^a \lambda^{a'} \lambda^{d_1}
\lambda^{d_2} \lambda^{b'} \lambda^b\right) &\equiv& - 4\, i\, g^4\, {\hat s 
\over |p'_{A\perp}|^2 |p'_{B\perp}|^2} C_{-+}(-p_B,p'_B)
A_{++}(k_1,k_2) \nonumber\\
A_{++}(k_1,k_2) &=& 2\, {{p'}_{A\perp}^* p'_{B\perp}\over k_{1\perp}}
{1\over k_{2\perp} - k_{1\perp} {k_2^+\over k_1^+}}\, ,\label{crb}
\end{eqnarray}
which has a massless divergence when gluons $1$ and $2$ become collinear.
As expected, the result is analogous to the one of eq.(\ref{tre}),
given the similarity of the color and the helicity structures. 
The multi-Regge limit, $y_1\gg y_2$, of eq.(\ref{crb}) is in agreement with 
the corresponding color ordering of eq.(\ref{fklh}), for $n=2$. 
We keep then fixed the position of gluons $A$ and $B$ 
in the color ordering and permute the outgoing gluons. The same considerations
made after eq.(\ref{tre}) apply here, namely the contribution of 
the ordering [$A,A',2,1,B',B$] (Fig.\ref{fig:centone}b) is obtained by 
exchanging the labels
$1$ and $2$ in eq.(\ref{crb}), however the ensuing coefficient is subleading
in the multi-Regge limit, $y_1\gg y_2$. Every other color configuration 
with all the gluons on the front of the lego plot is 
subleading to the required accuracy. 

\begin{figure}[hbt]
\vspace*{0cm}
\centerline{\epsfxsize=10cm \epsfbox{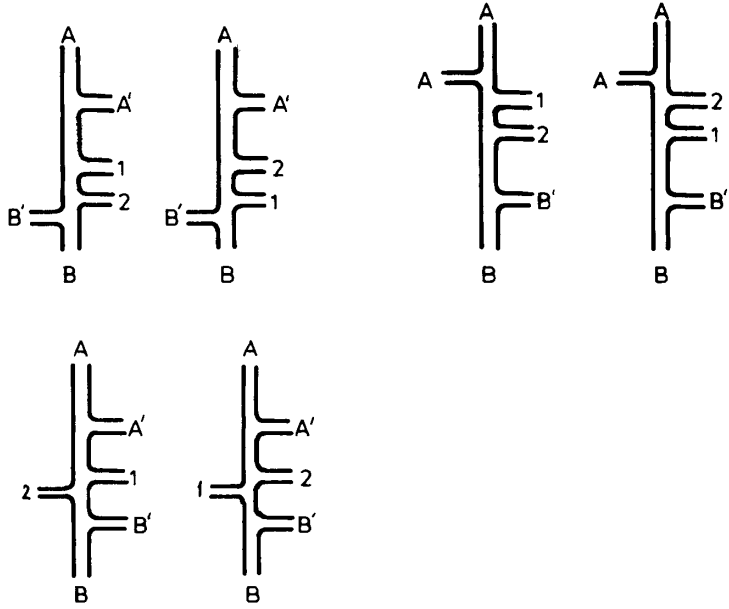}}
\vspace*{0.cm}
\caption{4-gluon production amplitude in the color orderings ($a$)
[$A,A',1,2,B,B'$] and ($b$) ($1\leftrightarrow 2$); ($c$) [$A,1,2,B',B,A'$] 
and ($d$) ($1\leftrightarrow 2$);
($e$) [$A,A',1,B',B,2$] and ($f$) ($1\leftrightarrow 2$).}
\label{fig:centdue}
\end{figure}

Then we move gluon $B$ one step to the left and consider the color orderings
\newline [$A,A',1,2,B,B'$] and [$A,1,2,B',B,A'$], and ($1\leftrightarrow 2$). 
These correspond respectively to have gluons $B'$ or $A'$ on the back of the
lego plot (Fig.\ref{fig:centdue}a, b, c, d); the respective strings of
spinor products may be related to eq.(\ref{crb}) through,
\begin{equation}
\langle p'_A k_1\rangle \langle k_2 p_B\rangle \langle p'_B p_A\rangle =
\langle p_A k_1\rangle \langle k_2 p'_B\rangle \langle p_B p'_A\rangle 
= \langle p'_A k_1\rangle \langle k_2 p'_B\rangle \langle p_B p_A\rangle\, 
,\label{crc}
\end{equation}
and the identity (\ref{flip}) (Appendix \ref{sec:appa}). It turns out that
the coefficients of the color orderings of Fig.\ref{fig:centdue}a,c are
equal to eq.(\ref{crb}) but with opposite sign. Then we take gluon 
$2$ to the back of the two-sided plot (Fig.~\ref{fig:centdue}e) and consider 
the color ordering [$A,A',1,B',B,2$]. Similarly to eq.(\ref{tred}), the 
spinor products yield,
\begin{eqnarray}
{\rm coeff.\; of}\; {\rm tr}\left(\lambda^a \lambda^{a'} \lambda^{d_1}
\lambda^{b'} \lambda^b \lambda^{d_2} \right) &\equiv& 4\, i\, g^4\, {\hat s 
\over |p'_{A\perp}|^2 |p'_{B\perp}|^2}\, C_{-+}(-p_B,p'_B)\,
B_{++}(k_1,k_2) \nonumber\\
B_{++}(k_1,k_2) &=& 2\, {{p'}_{A\perp}^* p'_{B\perp}\over
k_{1\perp} k_{2\perp}}\, .\label{crd}
\end{eqnarray}
Again, the same considerations made after eq.(\ref{tred}) apply here,
namely eq.(\ref{crd}) has no massless divergence when gluons $1$ and $2$ 
become collinear; it is already in agreement with the corresponding
color ordering in the multi-Regge limit (\ref{fklh}), for $n=2$; it
is invariant under the exchange of gluons $1$ and $2$, thus the color ordering
[$A,A',2,B',B,1$] (Fig.\ref{fig:centdue}f) yields the same contribution as 
in eq.(\ref{crd}), and in particular,
\begin{equation}
B_{++}(k_1,k_2) = A_{++}(k_1,k_2) + A_{++}(k_2,k_1)\, .\label{bpp}
\end{equation}

\begin{figure}[hbt]
\vspace*{0cm}
\centerline{\epsfxsize=10cm \epsfbox{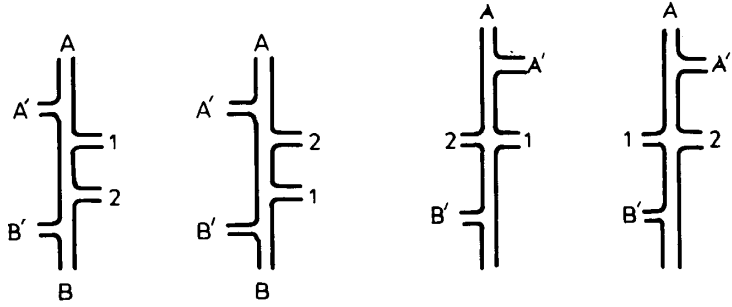}}
\vspace*{0.cm}
\caption{4-gluon production amplitude in the color orderings 
($a$) [$A,1,2,B,B',A'$] and ($b$) ($1\leftrightarrow 2$); ($c$)
[$A,A',1,B,B',2$] and ($d$) ($1\leftrightarrow 2$).}
\label{fig:centre}
\end{figure}

Then we move gluon $B$ one more step to the left, in order to have two
gluons on the front and two on the back of the lego plot, and
consider the color ordering [$A,1,2,B,B',A'$] (Fig. \ref{fig:centre}a)
and ($1\leftrightarrow 2$) (Fig. \ref{fig:centre}b), for
which the string of spinor products may be related to eq.(\ref{crb}) through,
\begin{equation}
\langle p_A k_1\rangle \langle k_2 p_B\rangle \langle p'_B p'_A\rangle 
= \langle p'_A k_1\rangle \langle k_2 p'_B\rangle \langle p_B p_A\rangle\,
.\label{cre}
\end{equation}
The color orderings [$A,A',1,B,B',2$] (Fig. \ref{fig:centre}c)
and ($1\leftrightarrow 2$) (Fig. \ref{fig:centre}d), are then obtained from
eq.(\ref{crd}), noticing that
\begin{equation}
\langle k_1 p_B\rangle \langle p_B p'_B\rangle \langle p'_B k_2\rangle = -
\langle k_1 p'_B\rangle \langle p'_B p_B\rangle \langle p_B k_2\rangle\,
.\label{crf}
\end{equation}
The other color orderings with two gluons on the front and two on the 
back of the lego plot are obtained by taking the color orderings
[$A,1,2,B,B',A'$], [$A,A',1,B,B',2$] and ($1\leftrightarrow 2$)
in reverse order, and by using the ciclicity of the traces and the
identity (\ref{flip}) (Appendix \ref{sec:appa}), which 
yields the same result as in eq.(\ref{cre}) and (\ref{crf}). Analogously,
the color orderings with three or four gluons on the back of the lego plot
are obtained by taking the color orderings [$A,A',1,2,B',B$],
[$A,A',1,2,B,B'$], [$A,1,2,B',B,A'$], [$A,A',1,B',B,2$] and 
($1\leftrightarrow 2$) in reverse order.

Substituting then equations (\ref{crb}-\ref{crf}), multiplied by the 
appropriate color orderings of
eq.(\ref{one}), we obtain the 4-gluon production amplitude in the 
helicity configuration (\ref{cra}) in the form of eq.(\ref{centr}),
with $C_{-+}(-p_A,p'_A)$, $C_{-+}(-p_B,p'_B)$, $A_{++}(k_1,k_2)$ and 
$B_{++}(k_1,k_2)$ given in eq.(\ref{centrc}), (\ref{crb})
and (\ref{crd}).
In the multi-Regge limit, $y_1\gg y_2$, the 8 color configurations obtained
by exchanging gluons $k_1$ and $k_2$ on the same side of the lego plot
(Fig.\ref{fig:centone}b, \ref{fig:centdue}b, \ref{fig:centdue}d, 
\ref{fig:centre}b and analogous with front and back of the
lego plot exchanged)
become subleading. Thus out of the 24 color configurations of eq.(\ref{centr}),
only the 16 configurations given by eq.(\ref{fpro}) for $n=2$ survive, and
eq.(\ref{centr}) is reduced to eq.(\ref{fklh}).

The other helicity configurations of eq.(\ref{crneg}) (Fig.~\ref{fig:cent}b, c,
d) are obtained from eq.(\ref{cra}), by substituting the 
product $\langle p_A p_B\rangle^4$ with the suitable 
product according to the pair of negative-helicity gluons considered. This 
entails to take the suitable complex conjugates of the $C$-vertices 
(\ref{centrc}), and the outcome is the same as in eq.(\ref{fklhel}) and
(\ref{fklhelb}). Inverting the helicity of all the gluons, and taking the
complex conjugates of the spinor products, we
obtain the configurations with two positive-helicity gluons. In particular,
we are interested in the configuration with gluons $k_1$ and $k_2$ having
negative helicity. To that purpose, we first compute the amplitude for the
configuration of Fig.~\ref{fig:cent}d,
\begin{eqnarray}
& & M(-p_A,+; p'_A,-; k_1,+; k_2,+; p'_B,-; -p_B,+) \label{centd}\\
& & = \left( p'_{B\perp}\over {p'}^*_{B\perp} \right)^2
M(-p_A,-; p'_A,+; k_1,+; k_2,+; p'_B,+; -p_B,-)\, ,\nonumber
\end{eqnarray}
and then we reverse all the helicities (Fig.~\ref{fig:cent}e), i.e. we
take in eq.(\ref{centd}) the complex conjugates of the spinor products. Thus
we obtain the amplitude for the configuration
$(-p_A,-; p'_A,+; k_1,-; k_2,-; p'_B,+; -p_B,-)$ in the form of
eq.(\ref{centr}), with $C_{-+}(-p_A,p'_A)$ and $C_{-+}(-p_B,p'_B)$ given in
eq.(\ref{centrc}), and $A_{--}(k_1,k_2) = A_{++}^*(k_1,k_2)$ and
$B_{--}(k_1,k_2) = B_{++}^*(k_1,k_2)$ derived from eq.(\ref{crb}) and 
(\ref{crd}). By taking then the multi-Regge limit of eq.(\ref{centr})
with $\nu_1=\nu_2=+$ and $\nu_1=\nu_2=-$, we obtain
\begin{eqnarray}
& & \lim_{y_1\gg y_2}\, M(-p_A,-; p'_A,+; k_1,-; k_2,-; p'_B,+; -p_B,-) 
\label{dflip}\\ & & = {p'_{A\perp} k_{1\perp} k_{2\perp} {p'}_{B\perp}^*\over 
{p'}_{A\perp}^* k_{1\perp}^* k_{2\perp}^* p'_{B\perp}}\, \lim_{y_1\gg y_2}\, 
M(-p_A,-; p'_A,+; k_1,+; k_2,+; p'_B,+; -p_B,-)\, ,\nonumber
\end{eqnarray}
in agreement with eq.(\ref{chh}), iterated two times.
Again, the other helicity configurations of eq.(\ref{crneg}) 
are obtained by substituting the 
product $[p_A p_B]^4$ with the suitable 
product according to the pair of positive-helicity gluons considered, and 
the outcome is the same as in eq.(\ref{fklhel}) and (\ref{fklhelb}).

In an analogous way to the generalization of the kinematics (\ref{qmreg}) 
to the production of $n+2$ gluons, with 2 gluons in a forward-rapidity 
region, done at the end of sect.~\ref{sec:treone}, 
we may generalize the amplitude (\ref{centr}) in the helicity configuration
of Fig.~\ref{fig:cent}a to the production of $n+2$ gluons, with 2 gluons with 
likewise rapidity in the central-rapidity region. As in sect.~\ref{sec:treone},
the color counting yields $3! 2^n$ leading color orderings, which for $n=2$ 
gives the 24 orderings considered in eq.(\ref{centr}).

\subsection{The amplitudes with 3 negative-helicity gluons}
\label{sec:fourbis}

The other helicity configurations we are interested in are the ones where
gluons $k_1$ and $k_2$ have opposite helicities (Fig.~\ref{fig:cent}f, g). 
In the multi-Regge kinematics, and in the next-to-leading corrections to it 
(\ref{crreg}), helicity is conserved in the forward-rapidity
regions (\ref{crneg}) and two of the gluons emitted there must have
negative helicity. Thus we need the subamplitudes in eq.(\ref{one}) 
to have 3 negative and 3 positive-helicity gluons. These
have been computed in ref.~\cite{bg}, \cite{mpz}, and are given in terms
of 3 inequivalent helicity orderings: $(++-+--)$,
$(+-+-+-)$ and $(+++---)$, singled out according to the color ordering. 
The subamplitudes may be written as \cite{mpz},
\begin{eqnarray}
i\, m(p_1; p_2; p_3; p_4; p_5; p_6) &=&
8\, i\, g^4\, \left({\alpha^2\over \hat t_{123} \hat s_{12} \hat s_{23}
\hat s_{45} \hat s_{56}} + {\beta^2\over \hat t_{234} \hat s_{23} \hat s_{34} 
\hat s_{56} \hat s_{61}}
\right. \nonumber\\ &+& \left. {\delta^2\over \hat t_{345} \hat s_{34}
\hat s_{45} \hat s_{61} \hat s_{12}} + {\hat t_{123}\, \beta\delta + 
\hat t_{234}\, \delta\alpha + \hat t_{345}\, \alpha\beta \over \hat s_{12} 
\hat s_{23} \hat s_{34} \hat s_{45} \hat s_{56} \hat s_{61}} \right)\, 
,\label{man}\\ {\rm with} & & \hat t_{ijk} = (p_i + p_j + p_k)^2 = 
\hat s_{ij} + \hat s_{jk} + \hat s_{ik}\, .\nonumber
\end{eqnarray}
The coefficients $\alpha$, $\beta$ and $\delta$ for the different
helicity orderings are,
\begin{equation} \begin{tabular}{cccc}
& $1^+ 2^+ 3^+ 4^- 5^- 6^-$ & $1^+ 2^+ 3^- 4^+ 5^- 6^-$ & $1^+ 2^- 3^+ 4^- 
5^+ 6^-$ \\ \hline
$\alpha$ & 0 & $-[12] \langle 56 \rangle \langle 4+|\gamma\cdot K|3+\rangle$ &
$[13] \langle 46 \rangle \langle 5+|\gamma\cdot K|2+\rangle$ \\
$\beta$ & $[23] \langle 56 \rangle \langle 1+|\gamma\cdot K|4+\rangle$ &
$[24] \langle 56 \rangle \langle 1+|\gamma\cdot K|3+\rangle$ &
$[51] \langle 24 \rangle \langle 3+|\gamma\cdot K|6+\rangle$ \\
$\delta$ & $[12] \langle 45 \rangle \langle 3+|\gamma\cdot K|6+\rangle$ &
$[12] \langle 35 \rangle \langle 4+|\gamma\cdot K|6+\rangle$ &
$[35] \langle 62 \rangle \langle 1+|\gamma\cdot K|4+\rangle$ \\ \hline
\multicolumn{4}{c}{$K=p_l + p_m + p_n$, with $l,m,n$ the gluons with  
positive helicity.} \end{tabular} \label{tabel}
\end{equation}

We are interested in the configuration 
$(-p_A,-; p'_A,+; k_1,+; k_2,-; p'_B,+; -p_B,-)$. We begin to examine it
in the color ordering [$A,A',1,2,B',B$] (Fig.~\ref{fig:centone}a), which is 
in the helicity ordering
$(p'_A,+; k_1,+; k_2,-; p'_B,+; -p_B,-; -p_A,-)$. Then the subamplitude 
(\ref{man}) has the form,
\begin{eqnarray}
& & {\rm coeff.\; of}\; {\rm tr}\left(\lambda^a \lambda^{a'} \lambda^{d_1}
\lambda^{d_2} \lambda^{b'} \lambda^b\right) \equiv - 4\, i\, g^4\, {\hat s 
\over |p'_{A\perp}|^2 |p'_{B\perp}|^2}\, C_{-+}(-p_B,p'_B)\,
A_{+-}(k_1,k_2) \nonumber\\ & &
= 8\, i\, g^4\, \left({\alpha^2\over \hat t_{A'12} \hat s_{A'1} \hat s_{12}
\hat s_{B'B} \hat s_{BA}} + {\beta^2\over \hat t_{12B'} \hat s_{12}
\hat s_{2B'} \hat s_{BA} \hat s_{AA'}} \right.\label{kosa}\\ & & \left. 
+ {\delta^2\over \hat t_{2B'B} \hat s_{2B'} \hat s_{B'B} \hat s_{AA'} 
\hat s_{A'1}}  + {\hat t_{A'12}\, \beta\delta + \hat t_{12B'}\, \delta\alpha + 
\hat t_{2B'B}\, \alpha\beta \over \hat s_{A'1} \hat s_{12} \hat s_{2B'} 
\hat s_{B'B} \hat s_{BA} \hat s_{AA'}} \right)\, ,\nonumber
\end{eqnarray}
with the $C$-vertex given in eq.(\ref{centrc}), the Mandelstam invariants 
taken from eq.(\ref{crinv}) (Appendix \ref{sec:appd}), and
\begin{eqnarray}
\alpha &=& - [p'_A k_1]\, \langle p_B p_A\rangle\, \langle p'_B+| \gamma\cdot 
(p'_A+k_1+p'_B) |k_2+\rangle \nonumber\\ 
\beta &=& [k_1 p'_B]\, \langle p_B p_A\rangle\, \langle p'_A+| \gamma\cdot 
(p'_A+k_1+p'_B) |k_2+\rangle \label{kosb}\\
\delta &=& [p'_A k_1]\, \langle k_2 p_B\rangle\, \langle p'_B+| \gamma\cdot 
(p'_A+k_1+p'_B) |p_A+\rangle\, \nonumber
\end{eqnarray}
from eq.(\ref{tabel}).
Using the spinor products (\ref{compspi}) (Appendix \ref{sec:appa}) and
(\ref{crpro}) (Appendix \ref{sec:appd}), and
the identity (\ref{flips}) (Appendix \ref{sec:appa}),
we compute the coefficients (\ref{kosb}) and the invariants
$\hat t_{ijk}$ in the kinematics of eq.(\ref{crreg}). In particular,
$\hat t_{2B'B}$ is
\begin{equation}
\hat t \equiv \hat t_{2B'B} = (k_2+p'_B-p_B)^2 = 
\hat t_{1AA'} \simeq - 
\left(|p'_{B\perp}+k_{2\perp}|^2+k_1^- k_2^+\right)\, .\label{tbb}
\end{equation}
Thus the vertex $A_{-+}$ in eq.(\ref{kosa}) becomes,
\begin{eqnarray}
A_{+-}(k_1,k_2) &=& -2\, {k_{1\perp}^* \over k_{1\perp}} 
\left\{ - {1\over \hat s_{12}} \left[{k_{2\perp}^2 |q_{A\perp}|^2 \over 
(k_1^-+k_2^-)k_2^+} +{k_{1\perp}^2 |q_{B\perp}|^2 \over (k_1^++k_2^+)k_1^-} 
+ {\hat t\, k_{1\perp}k_{2\perp}\over k_1^-k_2^+} \right]\right. \nonumber\\
&+& \left. {(q_{B\perp}+k_{2\perp})^2 \over \hat t} -
{q_{B\perp}+k_{2\perp} \over \hat s_{12}}
\left[{k_1^-+k_2^-\over k_1^-} k_{1\perp} - {k_1^++k_2^+\over k_2^+} 
k_{2\perp} \right]\right\} \label{kosc}\\
{\rm with} & & q_A = -(p'_A - p_A) \qquad q_B = p'_B - p_B\, .\nonumber
\end{eqnarray}
Note that besides the usual pole as gluons 1 and 2 become collinear,
eq.(\ref{kosc}) has a 3-particle pole as $\hat t\rightarrow 0$, i.e.
the amplitude factorizes into two subamplitudes connected by the propagator 
of the gluon exchanged between gluons 1 and 2. To obtain the coefficient of 
the color ordering [$A,A',2,1,B',B$] (Fig.~\ref{fig:centone}b), we must take 
the helicity ordering
$(-p_B,-; -p_A,-; p'_A,+; k_2,-; k_1,+; p'_B,+)$. This is obtained from
\newline
$(-p_B,+; -p_A,+; p'_A,-; k_2,+; k_1,-; p'_B,-)$ by 
reversing the helicities, i.e. by taking the complex 
conjugates of the corresponding spinor products, through the identities 
(\ref{flips}) and (\ref{flipc}) (Appendix \ref{sec:appa}). As we see from
the direct calculation, it amounts to exchange the labels 1 and 2 and
take the complex conjugate in eq.(\ref{kosc}),
\begin{equation}
{\rm coeff.\; of}\; {\rm tr}\left(\lambda^a \lambda^{a'} \lambda^{d_2}
\lambda^{d_1} \lambda^{b'} \lambda^b\right) = - 4\, i\, g^4\, {\hat s 
\over |p'_{A\perp}|^2 |p'_{B\perp}|^2}\, C_{-+}(-p_B,p'_B)\,
A_{+-}^*(k_2,k_1)\, .\label{unodue}
\end{equation}

In order to
compute the coefficient of the color ordering [$A,A',1,2,B,B'$], obtained 
by taking gluon $B'$ to the back of lego plot (Fig.~\ref{fig:centdue}a), 
we must consider the
helicity ordering $(k_2,-; -p_B,-; p'_B,+; -p_A,-; p'_A,+; k_1,+)$, 
which is obtained from \newline
$(k_2,+; -p_B,+; p'_B,-; -p_A,+; p'_A,-; k_1,-)$ by 
reversing the helicities. For the color ordering [$A,1,2,B',B,A'$]
(Fig.~\ref{fig:centdue}c), we must take the helicity ordering \newline
$(p'_A,+; -p_A,-; k_1,+; k_2,-; p'_B,+; -p_B,-)$ and $\alpha$,
$\beta$ and $\delta$ from the third column of eq.(\ref{tabel}); while for 
the color ordering [$A,1,2,B,B',A'$] (Fig.~\ref{fig:centre}a) 
we consider the helicity ordering
$(p'_B,+; p'_A,+; -p_A,-; k_1,+; k_2,-; -p_B,-)$.
To the required accuracy we obtain,
\begin{eqnarray}
& & {\rm coeff.\; of}\; {\rm tr}\left(\lambda^a \lambda^{a'} \lambda^{d_1}
\lambda^{d_2} \lambda^b \lambda^{b'}\right) =
{\rm coeff.\; of}\; {\rm tr}\left(\lambda^a \lambda^{d_1}
\lambda^{d_2} \lambda^{b'} \lambda^b \lambda^{a'}\right) \label{kosd}\\
& & = - {\rm coeff.\; of}\; {\rm tr}\left(\lambda^a \lambda^{d_1}
\lambda^{d_2} \lambda^b \lambda^{b'} \lambda^{a'}\right) = -
{\rm coeff.\; of}\; {\rm tr}\left(\lambda^a \lambda^{a'} \lambda^{d_1}
\lambda^{d_2} \lambda^{b'} \lambda^b\right)\, .\nonumber
\end{eqnarray}
For the color ordering [$A,A',1,B',B,2$], obtained by taking gluon $2$ to 
the back of lego plot (Fig.~\ref{fig:centdue}e), we take the helicity ordering 
$(p'_A,+; k_1,+; p'_B,+; -p_B,-; k_2,-; -p_A,-)$, for which the subamplitude
has the simpler form,
\begin{eqnarray}
& & {\rm coeff.\; of}\; {\rm tr}\left(\lambda^a \lambda^{a'} \lambda^{d_1}
\lambda^{b'} \lambda^b \lambda^{d_2} \right) \equiv 4\, i\, g^4\, {\hat s 
\over |p'_{A\perp}|^2 |p'_{B\perp}|^2}\, C_{-+}(-p_B,p'_B)\, B_{+-}(k_1,k_2) 
\label{kosz}\\ & & = 8\, i\, g^4\, \left({\beta^2\over \hat t_{1BB'} 
\hat s_{1B'} \hat s_{B'B} \hat s_{2A} \hat s_{AA'}} 
+ {\delta^2\over \hat t_{B'B2} \hat s_{B'B} \hat s_{B2} \hat s_{AA'} 
\hat s_{A'1}} + {\hat t_{A'1B'}\, \beta\delta \over \hat s_{A'1} \hat s_{1B'} 
\hat s_{B'B} \hat s_{B2} \hat s_{2A} \hat s_{AA'}} \right)\, ,\nonumber\\
\medskip & & {\rm with} \quad \hat t_{1B'B} = (k_1+p'_B-p_B)^2 = 
\hat t_{2AA'} = \simeq - \left(|p'_{B\perp}+k_{1\perp}|^2+k_1^+ k_2^-\right)\, 
,\nonumber
\end{eqnarray}
and $\beta$ and $\delta$ given by the first column of eq.(\ref{tabel}).
After substitution of the spinor products and of the Mandelstam invariants
we obtain,
\begin{equation}
B_{+-}(k_1,k_2) = - 2\,\left[ {k_{2\perp} (q_{B\perp}^*+k_{1\perp}^*)^2
\over k_{2\perp}^* \hat t_{1B'B}} + {k_{1\perp}^* (q_{B\perp}+k_{2\perp})^2
\over k_{1\perp} \hat t_{2B'B}} + {(q_{B\perp}^*+k_{1\perp}^*)
(q_{B\perp}+k_{2\perp}) \over k_{1\perp} k_{2\perp}^*}\right]\, .\label{kose} 
\end{equation}
Analogously to eq.(\ref{crd}), eq.(\ref{kose}) has no massless divergence 
when gluons $1$ and $2$ become collinear, since they are on opposite sides
of the lego plot. In addition, it is invariant under the exchange of the
labels 1 and 2 and the complex conjugation, and one can see that
\begin{equation}
B_{+-}(k_1,k_2) = A_{+-}(k_1,k_2) + A_{+-}^*(k_2,k_1)\, ,\label{cc}
\end{equation}
with $A_{+-}(k_1,k_2)$ given in eq.(\ref{kosc}). Finally, for the color 
ordering [$A,A',1,B,B',2$] (Fig.~\ref{fig:centre}c) we take the helicity 
ordering 
$(p'_A,+; k_1,+; -p_B,-; p'_B,+; k_2,-; -p_A,-)$ and eq.(\ref{man}) with
$\alpha$, $\beta$ and $\delta$ given by the second column of eq.(\ref{tabel}).
However, the additional 
terms which appear with respect to eq.(\ref{kosz}) are subleading to the 
required accuracy, and we obtain,
\begin{equation}
{\rm coeff.\; of}\; {\rm tr}\left(\lambda^a \lambda^{a'} \lambda^{d_1}
\lambda^b \lambda^{b'} \lambda^{d_2}\right) = -
{\rm coeff.\; of}\; {\rm tr}\left(\lambda^a \lambda^{a'} \lambda^{d_1}
\lambda^{b'} \lambda^b \lambda^{d_2}\right)\, .\label{kosf}
\end{equation}
The coefficients of all the color orderings other than the ones of 
eq.(\ref{kosa}), (\ref{kosd}), (\ref{kosz}) and (\ref{kosf}) are 
subleading, and collecting the results of eq.(\ref{kosa}) and 
(\ref{kosc}-\ref{kosf}) we obtain the amplitude for the configuration
$(-p_A,-; p'_A,+; k_1,+; k_2,-; p'_B,+; -p_B,-)$ in the form of 
eq.(\ref{centr}), with $C_{-+}(-p_A,p'_A)$, $C_{-+}(-p_B,p'_B)$,
$A_{+-}(k_1,k_2)$ and $B_{+-}(k_1,k_2)$ given in eq.(\ref{centrc}), 
(\ref{kosc}) and (\ref{kose}) respectively. In addition, for the color 
orderings
obtained by exchanging gluons 1 and 2 the related coefficients undergo the
complex conjugation, as shown in eq.(\ref{unodue}) and summarized by
eq.(\ref{exchange}). By taking then 
the multi-Regge limit of eq.(\ref{centr}) with $\nu_1=\nu_2=+$
and $\nu_1=+$, $\nu_2=-$, we obtain
\begin{eqnarray}
& & \lim_{y_1\gg y_2}\, M(-p_A,-; p'_A,+; k_1,+; k_2,-; p'_B,+; -p_B,-) 
\label{ccflip}\\ & & = {(p'_{A\perp}+k_{1\perp}) k_{2\perp} {p'}_{B\perp}^*
\over ({p'}_{A\perp}^*+k_{1\perp}^*) k_{2\perp}^* p'_{B\perp}}\, 
\lim_{y_1\gg y_2}\, M(-p_A,-; p'_A,+; k_1,+; k_2,+; p'_B,+; -p_B,-)\, 
,\nonumber
\end{eqnarray}
which, setting $p'_{A\perp}+k_{1\perp} = - q_{2\perp}$ and $p'_{B\perp}
= q_{3\perp}$, is in agreement with eq.(\ref{chh}).

The other configuration of interest is
$m(-p_A,-; p'_A,+; k_1,-; k_2,+; p'_B,+; -p_B,-)$. Using again eq.(\ref{man})
and (\ref{tabel}), we find that,
\begin{eqnarray}
& & {\rm coeff.\; of}\; {\rm tr}\left(\lambda^a \lambda^{a'} \lambda^{d_1}
\lambda^{d_2} \lambda^{b'} \lambda^b\right) = - 4\, i\, g^4\, {\hat s 
\over |p'_{A\perp}|^2 |p'_{B\perp}|^2}\, C_{-+}(-p_B,p'_B)\,
A_{+-}^*(k_1,k_2) \label{kosmp}\\
& & {\rm coeff.\; of}\; {\rm tr}\left(\lambda^a \lambda^{a'} \lambda^{d_1}
\lambda^{b'} \lambda^b \lambda^{d_2} \right) = 4\, i\, g^4\, {\hat s 
\over |p'_{A\perp}|^2 |p'_{B\perp}|^2}\, C_{-+}(-p_B,p'_B)\, 
B_{+-}^*(k_1,k_2)\, ,\nonumber 
\end{eqnarray}
with $A_{+-}(k_1,k_2)$ and $B_{+-}(k_1,k_2)$ defined in eq.(\ref{kosc}) and
(\ref{kose}), respectively. Accordingly the other leading color orderings
may be computed, and we obtain the amplitude for the configuration
$(-p_A,-; p'_A,+; k_1,-; k_2,+; p'_B,+; -p_B,-)$ in the form of 
eq.(\ref{centr}), with $A_{-+}(k_1,k_2)=A_{+-}^*(k_1,k_2)$ and
$B_{-+}(k_1,k_2)=B_{+-}^*(k_1,k_2)$.

In this section we have explicitly considered the amplitudes in the 
configurations $(-p_A,-; p'_A,+; k_1,\pm; k_2,\mp; p'_B,+; -p_B,-)$. 
Accordingly, we may reverse the helicities of the gluons produced in the 
forward-rapidity regions, as in eq.(\ref{crneg}). The effect of the reversal 
is the same as in eq.(\ref{fklhel}) and (\ref{fklhelb}), i.e. it amounts
to take the complex conjugate of the $C$-vertices in eq.(\ref{centrc}).
Thus each of the 16 leading helicity configurations of the 4-gluon
production amplitude in the kinematics (\ref{crreg}) may be cast in the
form of eq.(\ref{centr}).

\subsection{The Fadin-Lipatov amplitudes}
\label{sec:fourdue}

The 4-gluon production amplitude, with gluons $k_1$ and $k_2$ produced
in the central-rapidity region, according to the kinematics 
(\ref{crreg}), has been computed in ref.~\cite{fl}, and for a generic
helicity configuration is given by,
\begin{eqnarray}
& & i\, M^{aa'd_1d_2b'b}_{\nu_A\nu'_A\nu_1\nu_2\nu'_B\nu_B} = 
2\, i\, g^4\, {\hat s\over\hat s_{AA'}\hat s_{BB'}} f^{a'ca} f^{b'c'b}\,
\epsilon_{\mu_A}^{\nu_A*}(p_A) \epsilon_{\mu_B}^{\nu_B*}(p_B)
\epsilon_{\mu'_A}^{\nu'_A}(p'_A) \epsilon_{\mu'_B}^{\nu'_B}(p'_B)
\label{fad}\\ & & \times
\Gamma^{\mu_A\mu'_A}(p_A,p'_A,p_B)\, \Gamma^{\mu_B\mu'_B}(p_B,p'_B,p_A)\,
\left[\epsilon_{\mu_1}^{\nu_1}(k_1) \epsilon_{\mu_2}^{\nu_2}(k_2)
f^{cd_1j} f^{jd_2c'} A_{\mu_1\mu_2} + \left(\begin{array}{c} 
k_1\leftrightarrow k_2\\ \nu_1\leftrightarrow \nu_2\\ d_1\leftrightarrow d_2 
\end{array}\right) \right]\, ,\nonumber
\end{eqnarray}
where the $\Gamma$-tensors are given in eq.(\ref{gamm}), the Mandelstam 
invariants in eq.(\ref{crinv}) (Appendix \ref{sec:appd}), and the 2-gluon
production vertex $A$ in the central-rapidity region is\footnote{
Eq.(\ref{avert}) is easily derived from the expression of $A$ in 
ref.~\cite{fl} by using the Mandelstam invariants (\ref{crinv}) (Appendix 
\ref{sec:appd}), or from the one in ref.~\cite{lipatov} by writing the
Sudakov variables in terms of the light-cone ones, $\beta_i=k_i^+/p_A^+$
and $\alpha_i=k_i^-/p_B^-$.}
\begin{eqnarray}
& & A^{\mu_1\mu_2}(k_1,k_2) = -{a_1^{\mu_1}a_2^{\mu_2}\over \hat t} +
{b_1^{\mu_1}b_2^{\mu_2}\over\hat s} \left(1+{k_1^- k_2^+\over\hat t}\right) - 
{b_1^{\mu_1}c_2^{\mu_2}\over\hat s} \left({{p'}_B^+ p_A^+\over k_2^+ 
(k_1^+ + k_2^+)} + {p_A^+ k_1^- k_2^-\over p_B^-\hat t}\right) \nonumber\\
& & - {c_1^{\mu_1}b_2^{\mu_2}\over\hat s} \left({{p'}_A^- p_B^-\over k_1^- 
(k_1^- + k_2^-)} + {p_B^- k_1^+ k_2^+\over p_A^+\hat t}\right) -
{c_1^{\mu_1}c_2^{\mu_2}\over\hat s} \left(1 + {\hat s_{12}\over\hat t} -
{k_1^+ k_2^-\over\hat t}\right) \label{avert}\\ & & -2\left(g^{\mu_1\mu_2}
- {2k_2^{\mu_1}k_1^{\mu_2}\over \hat s_{12}}\right) \nonumber\\ & & \times
\left(1 + {\hat t\over\hat s_{12}} +
{k_1^-k_2^+\over \hat t} + {k_1^-k_2^+ - k_1^+ k_2^-\over\hat s_{12}} +
{k_2^-\over k_1^- + k_2^-} {p_A^+{p'}_A^-\over\hat s_{12}} +
{k_1^+\over k_1^+ + k_2^+} {p_B^-{p'}_B^+\over\hat s_{12}}\right)\, ,\nonumber
\end{eqnarray}
with
\begin{eqnarray} 
a_1 &=& -2\left[-q_A + \left({k_1^+\over p_A^+} - {{p'}_A^-\over k_1^-}\right)
p_A - {k_1^-\over p_B^-} p_B - {\hat t\over \hat s_{12}} k_2\right] \nonumber\\
a_2 &=& -2\left[q_B + \left({k_2^-\over p_B^-} - {{p'}_B^+\over k_2^+}\right)
p_B - {k_2^+\over p_A^+} p_A - {\hat t\over \hat s_{12}} k_1\right] \nonumber\\
b_1 &=& 2\left(p_B - {k_1^+ p_B^-\over\hat s_{12}}k_2\right) \qquad
b_2 = 2\left(p_A - {p_A^+ k_2^-\over\hat s_{12}}k_1\right) \label{adef}\\
c_1 &=& 2\left(p_A - {p_A^+ k_1^-\over\hat s_{12}}k_2\right) \qquad
c_2 = 2\left(p_B - {k_2^+ p_B^-\over\hat s_{12}}k_1\right) \nonumber
\end{eqnarray}
and $\hat t$ and $q_{A,B}$ defined in eq.(\ref{tbb}) and (\ref{kosc}). 
Rewriting the product of structure constants as a trace 
of nested commutators (\ref{fpro}), eq.(\ref{fad}) becomes
\begin{eqnarray}  
& & i\, M^{aa'd_1d_2b'b}_{\nu_A\nu'_A\nu_1\nu_2\nu'_B\nu_B} = -
4\, i\, g^4\, {\hat s\over\hat s_{AA'}\hat s_{BB'}} 
\epsilon_{\mu_A}^{\nu_A*}(p_A) \epsilon_{\mu_B}^{\nu_B*}(p_B)
\epsilon_{\mu'_A}^{\nu'_A}(p'_A) \epsilon_{\mu'_B}^{\nu'_B}(p'_B)
\epsilon_{\mu_1}^{\nu_1}(k_1) \epsilon_{\mu_2}^{\nu_2}(k_2)
\nonumber\\ & & \times
\Gamma^{\mu_A\mu'_A}(p_A,p'_A,p_B)\, \Gamma^{\mu_B\mu'_B}(p_B,p'_B,p_A)\,
\left\{ A_{\mu_1\mu_2}(k_1,k_2) 
\right. \label{fada}\\ & & \times
\left[ {\rm tr} \left( \lambda^a \lambda^{a'} \lambda^{d_1} \lambda^{d_2} 
\lambda^{b'} \lambda^b - \lambda^a \lambda^{a'} \lambda^{d_1} 
\lambda^{d_2} \lambda^b \lambda^{b'} - \lambda^a \lambda^{d_1} \lambda^{d_2}
\lambda^{b'} \lambda^b \lambda^{a'} + \lambda^a \lambda^{d_1} \lambda^{d_2}
\lambda^b \lambda^{b'} \lambda^{a'} \right) \right. \nonumber\\ & & \left.
+ {\rm traces}\, {\rm in}\, {\rm reverse}\, {\rm order} \right] 
- [A_{\mu_1\mu_2}(k_1,k_2)+A_{\mu_2\mu_1}(k_2,k_1)] \nonumber\\ & &
\left. \times \left[ {\rm tr} \left( 
\lambda^a \lambda^{a'} \lambda^{d_1} \lambda^{b'} \lambda^b \lambda^{d_2}
- \lambda^a \lambda^{a'} \lambda^{d_1} \lambda^b \lambda^{b'} \lambda^{d_2} 
\right) + {\rm traces}\, {\rm in}\, {\rm reverse}\, {\rm order} \right] 
+ (1 \leftrightarrow 2) \right\}\, .\nonumber
\end{eqnarray}
Eq.(\ref{fad}), or (\ref{fada}), describes 16 helicity configurations,
two for each of the produced gluons, since the helicity is conserved
by the $\Gamma$-tensors in the forward-rapidity regions. 
The gluon polarizations are chosen like 
in eq.(\ref{exh}) (Appendix \ref{sec:appa}), and the contractions of the 
$\Gamma$-tensors with the gluon polarizations are given by eq.(\ref{contra}).
Eq.(\ref{fada}) becomes,
\begin{eqnarray} 
& & i\, M^{aa'd_1d_2b'b}_{\nu_A\nu'_A\nu_1\nu_2\nu'_B\nu_B} 
\label{fadb}\\ & & = - 4\, i\, g^4\, 
{\hat s\over |p'_{A\perp}|^2 |p'_{B\perp}|^2}\, C_{\nu_A\nu'_A}(p_A,p'_A)\,
C_{\nu_B\nu'_B}(p_B,p'_B)\, \epsilon_{\mu_1}^{\nu_1}(k_1) 
\epsilon_{\mu_2}^{\nu_2}(k_2) \left\{ A_{\mu_1\mu_2}(k_1,k_2) \right. 
\nonumber\\ & & \times
\left[ {\rm tr} \left( \lambda^a \lambda^{a'} \lambda^{d_1} \lambda^{d_2} 
\lambda^{b'} \lambda^b - \lambda^a \lambda^{a'} \lambda^{d_1} 
\lambda^{d_2} \lambda^b \lambda^{b'} - \lambda^a \lambda^{d_1} \lambda^{d_2}
\lambda^{b'} \lambda^b \lambda^{a'} + \lambda^a \lambda^{d_1} \lambda^{d_2}
\lambda^b \lambda^{b'} \lambda^{a'} \right) \right. \nonumber\\ & & \left.
+ {\rm traces}\, {\rm in}\, {\rm reverse}\, {\rm order} \right] 
- [A_{\mu_1\mu_2}(k_1,k_2)+A_{\mu_2\mu_1}(k_2,k_1)] \nonumber\\ & &
\left. \times \left[ {\rm tr} \left( 
\lambda^a \lambda^{a'} \lambda^{d_1} \lambda^{b'} \lambda^b \lambda^{d_2}
- \lambda^a \lambda^{a'} \lambda^{d_1} \lambda^b \lambda^{b'} \lambda^{d_2} 
\right) + {\rm traces}\, {\rm in}\, {\rm reverse}\, {\rm order} \right] 
+ (1 \leftrightarrow 2) \right\}\, ,\nonumber
\end{eqnarray}
with the $C$-vertices defined like in eq.(\ref{centrc}),
\begin{eqnarray}
C_{++}(p_A,p'_A) &=& C_{--}(p_A,p'_A) = 1 \label{fadbb}\\
C_{++}(p_B,p'_B) &=& {{p'}_{B\perp}^* \over p'_{B\perp}} \qquad
C_{--}(p_B,p'_B) = C_{++}^*(p_B,p'_B)\, .\nonumber
\end{eqnarray}
The choice of reference vectors is arbitrary since the amplitude (\ref{fad})
is gauge invariant, so for the gluons $k_1$ and $k_2$ we choose the
polarizations $\epsilon_{\mu}^{\nu}(k_1,p_A)$ and 
$\epsilon_{\mu}^{\nu}(k_2,p_B)$. In the $A$-vertex (\ref{avert}) it is
convenient to contract first the vectors $a_1, b_1, c_1$ (\ref{adef}) with 
the gluon polarization $\epsilon_L^{\mu}(k_1)$ (\ref{lheap})
(Appendix \ref{sec:appa}), and the vectors $a_2, b_2, c_2$ with
$\epsilon_R^{\mu}(k_2)$, which yields the $A$-vertex in terms of transverse
polarizations \cite{lipatov}, and then to use the conversion 
(\ref{conv}) between the polarizations (\ref{exh}) and (\ref{lheap}). 
We begin with the $A$-vertex having helicities $\nu_1 = \nu_2 = +$. Using 
eq.(\ref{staa}) (Appendix \ref{sec:appa}), and after an appropriate
amount of algebra we obtain,
\begin{equation}
A_{\mu_1\mu_2}(k_1,k_2) \epsilon_{\mu_1}^+(k_1,p_A) \epsilon_{\mu_2}^+(k_2,p_B)
= A_{++}(k_1,k_2)\, ,\label{averpp}
\end{equation}
with $A_{++}(k_1,k_2)$ defined in eq.(\ref{crb}).
Setting $p'_{A\perp} = -q_{A\perp} \equiv -q_{1\perp}$ and
$p'_{B\perp} = q_{B\perp} \equiv q_{2\perp}$, and using the contraction
of the Lipatov vertex with the gluon polarizations (\ref{verc}), 
eq.(\ref{averpp}) fulfills the multi-Regge limit,
\begin{eqnarray}
\lim_{y_1\gg y_2} A_{\mu_1\mu_2} \epsilon_{\mu_1}^+(k_1,p_A) 
\epsilon_{\mu_2}^+(k_2,p_B) &=& - 2\, {q_{1\perp}^* q_{2\perp}\over
k_{1\perp} k_{2\perp}}\label{avermr}\\ &=& C(q_1,q_{12})\cdot\epsilon^+(k_1)\,
{1\over \hat t_{12}} C(q_{12},q_2)\cdot \epsilon^+(k_2)\, ,\nonumber
\end{eqnarray}
where $q_{12}$ is the momentum of the gluon exchanged between $k_1$ and
$k_2$ along the multi-Regge ladder, and $\hat t_{12} = - |q_{12\perp}|^2$.
Exchanging then gluons $k_1$ and $k_2$ in eq.(\ref{averpp}), we obtain
$A_{\mu_1\mu_2}(k_2,k_1) \epsilon_{\mu_1}^+(k_2) \epsilon_{\mu_2}^+(k_1)$,
which added to eq.(\ref{averpp}) gives,
\begin{equation}
[A_{\mu_1\mu_2}(k_1,k_2)+A_{\mu_2\mu_1}(k_2,k_1)]  \epsilon_{\mu_1}^+(k_1) 
\epsilon_{\mu_2}^+(k_2) = B_{++}(k_1,k_2)\, ,\label{averop}
\end{equation}
with $B_{++}(k_1,k_2)$ defined in eq.(\ref{crd}). 
Replacing eq.(\ref{averpp}) and
(\ref{averop}) into eq.(\ref{fadb}) we obtain the amplitude for the 
configuration $(p_A,\nu_A; p'_A,\nu'_A; k_1,+; k_2,+; p'_B,\nu'_B; p_B,\nu_B)$,
in agreement with its derivation eq.(\ref{centr}), with $\nu_1=\nu_2=+$,
from the PT amplitudes (sect.~\ref{sec:fourone}). Taking then the 
complex conjugates of eq.(\ref{averpp}) and (\ref{averop}) and substituting
them into eq.(\ref{fadb}), we obtain the amplitude in the configuration
$(p_A,\nu_A; p'_A,\nu'_A; k_1,-; k_2,-; p'_B,\nu'_B; p_B,\nu_B)$, in
agreement with eq.(\ref{centr}) with $\nu_1=\nu_2=-$.

The result of the contraction of the $A$-vertex with gluons $k_1$ and $k_2$
with opposite helicities is less simple. We obtain, after a bit of algebra,
\begin{equation}
A_{\mu_1\mu_2}(k_1,k_2) \epsilon_{\mu_1}^+(k_1,p_A) 
\epsilon_{\mu_2}^-(k_2,p_B) = A_{+-}(k_1,k_2)\, ,\label{averpm}
\end{equation}
with $A_{+-}(k_1,k_2)$ defined in eq.(\ref{kosc}). As a further check of
eq.(\ref{averpp}) and (\ref{averpm}), we note that the amplitude (\ref{fad}) 
must not diverge more rapidly than $1/|q_{i\perp}|$ in the collinear regions
$|q_{i\perp}|\rightarrow 0$, with $i=A,B$, in order for the related 
cross section not to diverge more than logarithmically \cite{fl}.
However for $|q_{i\perp}|\rightarrow 0$, the amplitude (\ref{fad}) has poles
due to the propagators $\hat s_{AA'} \simeq -|q_{A\perp}|^2$ and
$\hat s_{BB'} \simeq - |q_{B\perp}|^2$, which entails that the $A$-vertex
must be at least linear in $|q_{i\perp}|$,
\begin{equation}
\lim_{|q_{i\perp}|\rightarrow 0} A_{\mu_1\mu_2}(k_1,k_2) 
\epsilon_{\mu_1}^{\nu_1}(k_1,p_A) \epsilon_{\mu_2}^{\nu_2}(k_2,p_B) = 
O(|q_{i\perp}|) \label{alim}
\end{equation}
with $i=A,B$, which is fulfilled by eq.(\ref{averpp}) and (\ref{averpm}).

Exchanging the labels 1 and 2 and taking the complex conjugate in 
eq.(\ref{averpm}), we obtain $A_{\mu_1\mu_2}(k_2,k_1) 
\epsilon_{\mu_1}^-(k_2) \epsilon_{\mu_2}^+(k_1)$, which added to
eq.(\ref{averpm}) gives, after a bit of algebra,
\begin{equation}
[A_{\mu_1\mu_2}(k_1,k_2)+A_{\mu_2\mu_1}(k_2,k_1)]  \epsilon_{\mu_1}^+(k_1) 
\epsilon_{\mu_2}^-(k_2) = B_{+-}(k_1,k_2)\, ,\label{dpm}
\end{equation}
with $B_{+-}(k_1,k_2)$ defined in eq.(\ref{kose}). After substituting
eq.(\ref{averpm}) and (\ref{dpm}) into eq.(\ref{fadb}) we obtain
the amplitude in the configuration 
$(p_A,\nu_A; p'_A,\nu'_A; k_1,+; k_2,-; p'_B,\nu'_B; p_B,\nu_B)$, in agreement
with its derivation, eq.(\ref{centr}) with $\nu_1=+$ and $\nu_2=-$, from 
the helicity amplitudes with 3 negative-helicity gluons 
(sect.~\ref{sec:fourbis}). Taking then the complex conjugates of
eq.(\ref{averpm}) and (\ref{dpm}), we obtain the amplitude in the 
configuration \newline
$(p_A,\nu_A; p'_A,\nu'_A; k_1,-; k_2,+; p'_B,\nu'_B; p_B,\nu_B)$,
in agreement with eq.(\ref{kosmp}).

\section{Conclusions}
\label{sec:conc}

Using the helicity formalism, we have computed the real next-to-leading
corrections to the FKL amplitudes in the forward-rapidity region,
eq.(\ref{trepos}), and in the central-rapidity region, eq.(\ref{centr});
and we have shown that they are equivalent to the Fadin-Lipatov amplitudes,
eq.(\ref{five}) and (\ref{fad}) respectively, for the corresponding
helicity configurations.

We note that the algebra involved in deriving eq.(\ref{trepos})
and (\ref{centr}) is simpler starting from the helicity formalism than
from the general expression of the Fadin-Lipatov amplitudes,
eq.(\ref{five}) and (\ref{fad}). This hints that even the real corrections
beyond the next-to-leading order, which at the present time are known
only formally \cite{lipatov}, could be explicitly computed\footnote{A 
formalism to compute the amplitude for the production 
of $n+2$ gluons with a cluster of $m$ gluons, with $m\le n+1$, in the 
forward-rapidity or in the central-rapidity regions, has been considered
in ref.~\cite{lipatov}. We merely note here that the corresponding color 
structures may be easily inferred from eq.(\ref{fourn}) or from the
generalization of eq.(\ref{centr}), and the color counting
yields $(m+1)!\, 2^{n+2-m}$ leading color configurations.}.

In addition, the computation of the square of the vertices
for the production of two gluons with likewise rapidity in the 
forward-rapidity or in the central-rapidity regions \cite{progress},
which is needed to 
evaluate the real next-to-leading logarithmic corrections to the kernel of 
the BFKL equation, is of course simpler starting from the amplitudes 
at fixed helicities, eq.(\ref{trepos}) and (\ref{centr}), than from their
general expression, eq.(\ref{five}) and (\ref{fad}).

As remarked in the Introduction, in the next-to-leading corrections to the 
multi-Regge 
kinematics, and accordingly in the real next-to-leading logarithmic 
corrections to the kernel of the BFKL equation, also the production of
a quark-antiquark pair with likewise rapidity in the forward-rapidity or 
in the central-rapidity regions contributes \cite{fiore}, \cite{progress}.
The analysis of the corresponding amplitudes in the helicity formalism
is left for the future.

\section*{Acknowledgements}
 
I wish to thank Victor Fadin and Lev Lipatov for useful discussions.
 
\appendix
\section{Multiparton kinematics}
\label{sec:appa}

We consider the production of $n+2$ gluons of momentum $p_i$, with 
$i=0,...,n+1$ and $n\ge 0$, in the scattering between two gluons of momenta 
$p_A$ and $p_B$. Using light-cone coordinates $p^{\pm}= p_0\pm p_z$, and
complex transverse coordinates $p_{\perp} = p_x + i p_y$, with scalar
product $2 p\cdot q = p^+q^- + p^-q^+ - p_{\perp} q^*_{\perp} - p^*_{\perp} 
q_{\perp}$, the gluon 4-momenta are,
\begin{eqnarray}
p_A &=& \left(p_A^+, 0; 0, 0\right)\, ,\nonumber \\
p_B &=& \left(0, p_B^-; 0, 0\right)\, ,\label{in}\\
p_i &=& \left(|p_{i\perp}| e^{y_i}, |p_{i\perp}| e^{-y_i}; 
|p_{i\perp}|\cos{\phi_i}, |p_{i\perp}|\sin{\phi_i}\right)\, ,\nonumber
\end{eqnarray}
where to the left of the semicolon we have the + and -
components, and to the right the transverse components.
$y$ is the gluon rapidity and $\phi$ is the azimuthal angle between the 
vector $p_{\perp}$ and an arbitrary vector in the transverse plane.
From the momentum conservation,
\begin{eqnarray}
0 &=& \sum_{i=0}^{n+1} p_{i\perp}\, ,\nonumber \\
p_A^+ &=& \sum_{i=0}^{n+1} p_i^+\, ,\label{nkin}\\ 
p_B^- &=& \sum_{i=0}^{n+1} p_i^-\, ,\nonumber
\end{eqnarray}
the Mandelstam invariants may be written as,
\begin{eqnarray}
\hat s &=& 2 p_A\cdot p_B = \sum_{i,j=0}^{n+1} p_i^+ p_j^- \nonumber\\ 
\hat s_{Ai} &=& -2 p_A\cdot p_i = -\sum_{j=0}^{n+1} p_i^- p_j^+ \label{inv}\\ 
\hat s_{Bi} &=& -2 p_B\cdot p_i = -\sum_{j=0}^{n+1} p_i^+ p_j^- \nonumber\\ 
\hat s_{ij} &=& 2 p_i\cdot p_j = p_i^+ p_j^- + p_i^- p_j^+
- p_{i\perp} p_{j\perp}^* - p_{i\perp}^* p_{j\perp}\, .\nonumber
\end{eqnarray}

Massless Dirac spinors $\psi_{\pm}(p)$ of fixed helicity are
defined by the projection,
\begin{equation}
\psi_{\pm}(p) = {1\pm \gamma_5\over 2} \psi(p)\, ,\label{spi}
\end{equation}
with the shorthand notation,
\begin{eqnarray}
\psi_{\pm}(p) &=& |p\pm\rangle, \qquad \overline{\psi_{\pm}(p)} = 
\langle p\pm|\, ,\nonumber\\
\langle p k\rangle &=& \langle p- | k+ \rangle = \overline{\psi_-(p)}
\psi_+(k)\, ,\label{cpro}\\ 
\left[pk\right] &=& \langle p+ | k- \rangle = \overline{\psi_+(p)}\psi_-(k)\, 
.\nonumber
\end{eqnarray}
Using the spinor representation of ref. \cite{ptlip}, we can write
\begin{eqnarray}
\langle p_i p_j\rangle &=& p_{i\perp}\sqrt{p_j^+\over p_i^+} - p_{j\perp}
\sqrt{p_i^+\over p_j^+}\, ,\nonumber\\ \langle p_A p_i\rangle &=& -\sqrt{p_A^+
\over p_i^+}\, p_{i\perp}\, ,\label{spro}\\ \langle p_i p_B\rangle &=&
-\sqrt{p_B^- p_i^+}\, ,\nonumber\\ \langle p_A p_B\rangle 
&=& -\sqrt{\hat s}\, ,\nonumber
\end{eqnarray}
where we have used the mass-shell condition 
$|p_{i\perp}|^2 = p_i^+ p_i^-$. We consider also the spinor products
$\langle p_i+| \gamma\cdot p_k |p_j+\rangle$, which in the spinor 
representation of ref.~\cite{ptlip} take the form,
\begin{eqnarray}
\langle p_i+| \gamma\cdot p_k |p_j+\rangle &=& {1\over\sqrt{p_i^+ p_j^+}}
\left(p_i^+ p_j^+ p_k^- - p_i^+ p_{j\perp} p_{k\perp}^* - p_{i\perp}^* p_j^+
p_{k\perp} + p_{i\perp}^* p_{j\perp} p_k^+ \right) \nonumber\\
\langle p_i+| \gamma\cdot p_j |p_A+\rangle &=& \sqrt{p_A^+\over p_i^+}
\left(p_i^+ p_j^- - p_{i\perp}^* p_{j\perp}\right) \label{compspi}\\
\langle p_i+| \gamma\cdot p_j |p_B+\rangle &=& \sqrt{p_B^-\over p_i^+}
\left(-p_i^+ p_{j\perp}^* + p_{i\perp}^* p_j^+\right)\, .\nonumber
\end{eqnarray}
The spinor products fulfill the identities,
\begin{eqnarray}
\langle p k\rangle &=& - \langle k p\rangle \label{flip}\\
\langle p k\rangle^* &=& \left[kp\right] \label{flips}\\
\left( \langle p_i+| \gamma^{\mu}  |p_j+\rangle \right)^* &=&
\langle p_j+| \gamma^{\mu}  |p_i+\rangle \label{flipc}\\
\langle p k\rangle \left[kp\right] &=& 2p\cdot k = |\hat{s}_{pk}|\, .\nonumber
\end{eqnarray}

Throughout the paper the following representation for the gluon 
polarization is used \cite{mp},
\begin {equation}
\epsilon_{\mu}^{\pm}(p,k) = \pm {\langle p\pm |\gamma_{\mu}| k\pm\rangle\over
\sqrt{2} \langle k\mp | p\pm \rangle}\, ,\label{hpol}
\end{equation}
which enjoys the properties
\begin {eqnarray}
\epsilon_{\mu}^{\pm *}(p,k) &=& \epsilon_{\mu}^{\mp}(p,k)\, ,\nonumber\\
\epsilon_{\mu}^{\pm}(p,k)\cdot p &=& \epsilon_{\mu}^{\pm}(p,k)\cdot k = 0\,
,\label{polc}\\
\sum_{\nu=\pm} \epsilon_{\mu}^{\nu}(p,k) \epsilon_{\rho}^{\nu *}(p,k) &=&
- g_{\mu\rho} + {p_{\mu} k_{\rho} + p_{\rho} k_{\mu}\over p\cdot k}\, 
,\nonumber
\end{eqnarray}
where $k$ is an arbitrary light-like momentum. The sum in eq.(\ref{polc}) is
equivalent to use an axial, or physical, gauge. We use the explicit
representation of the gluon polarizations obtained in ref. \cite{ptlip},
\begin{eqnarray}
\epsilon_{\mu}^+(p_i, p_A) &=& -{p_{i\perp}^*\over p_{i\perp}}\, \left(
{\sqrt{2}\, p_{i\perp}\over p_i^-}, 0; {1\over\sqrt{2}}, {i\over\sqrt{2}}
\right)\, ,\nonumber\\ \epsilon_{\mu}^+(p_B, p_A) &=& - \left(0, 0; 
{1\over\sqrt{2}}, {i\over\sqrt{2}}\right) ,\label{exh}\\
\epsilon_{\mu}^+(p_A, p_B) &=& \left(0, 0; {1\over\sqrt{2}}, -{i\over
\sqrt{2}}\right)\, ,\nonumber\\ \epsilon_{\mu}^+(p_i, p_B) &=& \left(0,
{\sqrt{2}\, p_{i\perp}^*\over p_i^+}; {1\over\sqrt{2}}, -{i\over\sqrt{2}}
\right)\, ,\nonumber
\end{eqnarray}
in light-cone coordinates, and
introduce the {\sl left} and {\sl right} physical gauges \cite{lipat},
defined respectively by the conditions $\epsilon_L(p)\cdot p_A = 0$ and
$\epsilon_R(p)\cdot p_B = 0$. Accordingly, the decomposition of a 
polarization vector in light-cone or Sudakov components is,
\begin{eqnarray}
\epsilon_L^{\mu}(p) &=& \epsilon_{L\perp}^{\mu} - {p\cdot\epsilon_{L\perp}\over
p\cdot p_A}\, p_A^{\mu}\, ,\label{lheap}\\
\epsilon_R^{\mu}(p) &=& \epsilon_{R\perp}^{\mu} - {p\cdot\epsilon_{R\perp}\over
p\cdot p_B}\, p_B^{\mu}\, .\nonumber
\end{eqnarray}
The polarizations $\epsilon_L^{\mu}(p),\, \epsilon_R^{\mu}(p)$ for a momentum 
$p$ not in the beam direction are related by a gauge tranformation,
\begin{equation}
\epsilon_R^{\mu}(p) = \epsilon_L^{\mu}(p) + 2\, {\epsilon_{L,\perp}\cdot p\over
|p_{\perp}|^2}\, p^{\mu}\, ,\label{gaa}
\end{equation}
which for the transverse components $\epsilon_{L\perp}^{\mu},\, 
\epsilon_{R\perp}^{\mu}$ may be written as,
\begin{equation}
\epsilon_{R\perp}(p) = - {p_{\perp}\over p_{\perp}^*} \epsilon_{L\perp}^*(p)\,
.\label{gac}
\end{equation}
We impose then the following standard polarizations,
\begin{eqnarray}
\epsilon_{L\perp}^{\mu\pm}(p) &=& \left(0, 0; {1\over\sqrt{2}}, \pm {i\over
\sqrt{2}}\right)\, ,\label{staa}\\
\epsilon_{L\perp}^{\mu\pm}(p_B) &=& \epsilon_{R\perp}^{\mu\mp}(p_A) = \left(0, 
0; {1\over\sqrt{2}}, \pm {i\over\sqrt{2}}\right)\, .\label{stab}
\end{eqnarray}
We determine $\epsilon_L^{\mu}(p)$ using eq.(\ref{staa}) in the definition 
(\ref{lheap}), then we find $\epsilon_R^{\mu}(p)$ using the gauge 
transformations (\ref{gaa}) and (\ref{gac}) on $\epsilon_L^{\mu}(p)$.
Finally, $\epsilon_L^{\mu}(p_B)$ and $\epsilon_R^{\mu}(p_A)$ are given by
eq.(\ref{stab}) and the definitions (\ref{lheap}).
Comparing the results with eq.(\ref{exh}),
we obtain the following conversion table among the representations 
(\ref{hpol}) and (\ref{lheap}) of the polarizations
\begin{eqnarray}
\epsilon_{\mu}^+(p_i, p_A) &=& -{p_{i\perp}^*\over p_{i\perp}}\, 
\epsilon_{L\mu}^+(p_i)\, ,\nonumber\\ \epsilon_{\mu}^+(p_i, p_B) &=& 
-{p_{i\perp}^*\over p_{i\perp}}\, \epsilon_{R\mu}^+(p_i)\, \label{conv}\\ 
\epsilon_{\mu}^+(p_B, p_A) &=& -\epsilon_{L\mu}^+(p_B)\, ,\nonumber\\
\epsilon_{\mu}^+(p_A, p_B) &=& \epsilon_{R\mu}^+(p_A)\, .\nonumber
\end{eqnarray}

\section{Multi-Regge kinematics}
\label{sec:appb}

In the multi-Regge kinematics, we require that the gluons
are strongly ordered in rapidity and have comparable transverse momentum
(\ref{mreg}),
\begin{equation}
y_0 \gg y_1 \gg ...\gg y_{n+1};\qquad |p_{i\perp}|\simeq|p_{\perp}|\, 
.\nonumber
\end{equation}
Momentum conservation (\ref{nkin}) then becomes
\begin{eqnarray}
0 &=& \sum_{i=0}^{n+1} p_{i\perp}\, ,\nonumber \\
p_A^+ &\simeq& p_0^+\, ,\label{mrkin}\\ 
p_B^- &\simeq& p_{n+1}^-\, .\nonumber
\end{eqnarray}
The Mandelstam invariants (\ref{inv}) are reduced to,
\begin{eqnarray}
\hat s &=& 2 p_A\cdot p_B \simeq p_0^+ p_{n+1}^- \nonumber\\ 
\hat s_{Ai} &=& -2 p_A\cdot p_i \simeq - p_0^+ p_i^- \label{mrinv}\\ 
\hat s_{Bi} &=& -2 p_B\cdot p_i \simeq - p_i^+ p_{n+1}^- \nonumber\\ 
\hat s_{ij} &=& 2 p_i\cdot p_j \simeq |p_{i\perp}| |p_{j\perp}| e^{|y_i-y_j|}\,
,\nonumber
\end{eqnarray}
to leading accuracy. The spinor products (\ref{spro}) become,
\begin{eqnarray}
\langle p_i p_j\rangle &\simeq& -\sqrt{p_i^+\over p_j^+}\,
p_{j\perp}\, \qquad {\rm for}\, y_i>y_j ,\nonumber\\
\langle p_A p_i\rangle &\simeq& -\sqrt{p_0^+\over p_i^+}\,
p_{i\perp}\, ,\label{mrpro}\\ \langle p_i p_B\rangle 
&\simeq& -\sqrt{p_i^+ p_{n+1}^-}\, ,\nonumber\\ 
\langle p_A p_B\rangle &\simeq& -\sqrt{p_0^+ p_{n+1}^-}\, .\nonumber
\end{eqnarray}

\section{Next-to-leading corrections in the forward-rapidity region}
\label{sec:appc}

We consider the production of 3 gluons of momenta $k_1$, $k_2$ and $p'$
in the scattering between two gluons of momenta $k_0$ and $p$, with
gluon 4-momenta as given by eq.(\ref{in}), with $k_0\equiv p_A$ and
$p\equiv p_B$. Gluons $k_1$ and $k_2$ are produced in the forward-rapidity 
region of gluon $k_0$ with likewise rapidity, and are separated by a large
rapidity interval from $p'$. In addition,
the produced gluons have comparable transverse momenta (\ref{qmreg}),
\begin{equation}
y_1 \simeq y_2 \gg y'\,;\qquad |k_{1\perp}|\simeq|k_{2\perp}|\simeq|p'_{\perp}|
\, .\nonumber
\end{equation}
Momentum conservation (\ref{nkin}) then yields,
\begin{eqnarray}
0 &=& k_{1\perp} + k_{2\perp} + p'_{\perp}\, ,\nonumber \\
k_0^+ &\simeq& k_1^+ + k_2^+\, ,\label{frkin}\\ 
p^- &\simeq& p'^-\, ,\nonumber
\end{eqnarray}
and accordingly the Mandelstam invariants (\ref{inv}) may be written as,
\begin{eqnarray}
\hat s &=& 2 k_0\cdot p\, \simeq\, (k_1^+ + k_2^+) p'^-\, ,\nonumber \\
\hat u &=& -2 k_0\cdot p'\, \simeq\, - (k_1^+ + k_2^+) p'^-\, ,\nonumber \\
\hat u_i &=& -2 p\cdot k_i\, \simeq\, - k_i^+ p'^-\, ,\nonumber \\ \hat t_i 
&=& -2 k_0\cdot k_i\,\simeq\, - (k_1^+ + k_2^+) k_i^-\, ,\label{frinv}\\
\hat t &=& -2 p\cdot p'\, \simeq\, - |p'_{\perp}|^2\, ,\nonumber\\
\hat s_{12} &=& 2 k_1\cdot k_2 = k_1^+ k_2^- + k_1^- k_2^+
- k_{1\perp} k_{2\perp}^* - k_{1\perp}^* k_{2\perp}\, ,\nonumber
\end{eqnarray}
to leading accuracy, with $i=1,2$. The spinor products (\ref{spro}) become
\begin{eqnarray}
\langle k_0 p\rangle &=& -\sqrt{\hat s} \simeq - \sqrt{(k_1^+ + k_2^+) p'^-}\,
,\nonumber\\ 
\langle k_0 p'\rangle &=& -\sqrt{k_0^+\over p'^+}\, p'_{\perp} \simeq
{p'_{\perp}\over |p'_{\perp}|} \langle k_0 p\rangle\, ,\nonumber\\
\langle k_0 k_i\rangle &=& -\sqrt{k_0^+\over k_i^+}\, k_{i\perp}
\simeq -\sqrt{k_1^+ + k_2^+\over k_i^+} k_{i\perp}\, ,\nonumber\\
\langle k_i p\rangle &=& -\sqrt{p^- k_i^+}\, 
\simeq - \sqrt{k_i^+ p'^-}\, ,\label{frpro}\\
\langle p' p\rangle &=& -\sqrt{p^- p'^+}\,\simeq - |p'_{\perp}|\, ,\nonumber\\
\langle k_i p'\rangle &=& k_{i\perp}\sqrt{p'^+\over k_i^+} - p'_{\perp}
\sqrt{k_i^+\over p'^+} \simeq - p'_{\perp}\, \sqrt{k_i^+\over p'^+}\, 
,\nonumber \\
\langle k_1 k_2\rangle &=& k_{1\perp}\sqrt{k_2^+\over k_1^+} - 
k_{2\perp}\sqrt{k_1^+\over k_2^+}\, ,\nonumber
\end{eqnarray}
with $i=1,2$. From eq.(\ref{exh}), the gluon polarization vectors are
\begin{eqnarray}
\epsilon_{\mu}^+(p, k_0) &=& - \left(0, 0; 
{1\over\sqrt{2}}, {i\over\sqrt{2}}\right) \nonumber\\
\epsilon_{\mu}^+(k_0, p) &=& \left(0, 0; {1\over\sqrt{2}}, -{i\over
\sqrt{2}}\right) \label{frexh}\\
\epsilon_{\mu}^+(p_i, k_0) &=& -{p_{i\perp}^*\over p_{i\perp}}\, \left(
{\sqrt{2}\, p_{i\perp}\over p_i^-}, 0; {1\over\sqrt{2}}, {i\over\sqrt{2}}
\right) \nonumber\\  \epsilon_{\mu}^+(p_i, p) &=& \left(0,
{\sqrt{2}\, p_{i\perp}^*\over p_i^+}; {1\over\sqrt{2}}, -{i\over\sqrt{2}}
\right)\, ,\nonumber
\end{eqnarray}
with $p_i = k_1, k_2, p'$.

We straightforwardly extend the kinematics (\ref{qmreg}) to the production 
of $n+2$ gluons of momenta $k_1$, $k_2$ and $p_i$, with $i=1,...,n$, in the 
scattering between two gluons of momenta $k_0$ and $p$, with gluons 
$k_1$ and $k_2$ in the forward-rapidity region of gluon $k_0$,
\begin{equation}
y_{k_1} \simeq y_{k_2} \gg y_{p_1}\gg ...\gg y_{p_n}\,;\qquad |k_{1\perp}|
\simeq |k_{2\perp}| \simeq |p_{i\perp}|\, ,\label{qmrapp}
\end{equation}
with $1=i,...,n$. Momentum conservation (\ref{frkin}) simply generalizes to,
\begin{eqnarray}
0 &=& k_{1\perp} + k_{2\perp} + \sum_{i=1}^n p_{i\perp}\, ,\nonumber \\
k_0^+ &\simeq& k_1^+ + k_2^+\, ,\label{frkapp}\\ 
p^- &\simeq& p_n^-\, .\nonumber
\end{eqnarray}
In the spinor products (\ref{frpro}) we must replace the label
$p'$ with $p_n$, and consider the additional spinor products
\begin{eqnarray}
\langle k_0 p_i\rangle &=& -\sqrt{k_0^+\over p_i^+}\, p_{i\perp}
\simeq -\sqrt{k_1^+ + k_2^+\over p_i^+} p_{i\perp}\, ,\nonumber\\
\langle p_i p\rangle &=& -\sqrt{p^- p_i^+}\, 
\simeq - \sqrt{p_i^+ p_n^-}\, ,\label{frpapp}\\
\langle k_j p_i\rangle &=& k_{j\perp}\sqrt{p_i^+\over k_j^+} - p_{i\perp}
\sqrt{k_j^+\over p_i^+} \simeq - p_{i\perp}\, \sqrt{k_j^+\over p_i^+}\, 
,\nonumber
\end{eqnarray}
with $i=1,...,n-1$ and $j=1,2$.

\section{Next-to-leading corrections in the central-rapidity region}
\label{sec:appd}

We consider the production of 4 gluons of 
momenta $p'_A$, $k_1$, $k_2$ and $p'_B$ in the scattering between two gluons 
of momenta $p_A$ and $p_B$. We require that gluons $k_1$ and $k_2$ have 
likewise rapidity and are produced far away from the forward-rapidity regions,
with all the gluons having comparable transverse momenta,
\begin{equation}
y'_A\gg y_1 \simeq y_2 \gg y'_B\,;\qquad |k_{1\perp}| \simeq |k_{2\perp}|
\simeq |p'_{A\perp}| \simeq |p'_{B\perp}|\, .\nonumber
\end{equation}
The momentum conservation has the same form as in the multi-Regge kinematics
(\ref{mrkin})
\begin{eqnarray}
0 &=& p'_{A\perp} + k_{1\perp} + k_{2\perp} + p'_{B\perp} \nonumber \\
p_A^+ &\simeq& {p'}_A^+ \label{crkin}\\ 
p_B^- &\simeq& {p'}_B^-\, .\nonumber
\end{eqnarray}
To leading accuracy, and except for $\hat s_{12}$, the Mandelstam invariants
are the same as in the multi-Regge kinematics (\ref{mrinv}),
\begin{eqnarray}
\hat s &=& 2 p_A\cdot p_B \simeq {p'}_A^+ {p'}_B^- \nonumber\\ 
\hat s_{AA'} &=& -2 p_A\cdot p'_A \simeq -|p'_{A\perp}|^2 \nonumber\\
\hat s_{Ai} &=& -2 p_A\cdot k_i \simeq - {p'}_A^+ k_i^- \nonumber\\
\hat s_{AB'} &=& -2 p_A\cdot p'_B \simeq - {p'}_A^+ {p'}_B^- \nonumber\\
\hat s_{BA'} &=& -2 p_B\cdot p'_A \simeq - {p'}_A^+ {p'}_B^- \nonumber\\  
\hat s_{Bi} &=& -2 p_B\cdot k_i \simeq - k_i^+ {p'}_B^- \label{crinv}\\ 
\hat s_{BB'} &=& -2 p_B\cdot p'_B \simeq -|p'_{B\perp}|^2 \nonumber\\
\hat s_{A'i} &=& 2 p'_A\cdot k_i \simeq {p'}_A^+ k_i^- \nonumber\\
\hat s_{B'i} &=& 2 p'_B\cdot k_i \simeq k_i^+ {p'}_B^- \nonumber\\
\hat s_{A'B'} &=& 2 p'_A\cdot p'_B \simeq {p'}_A^+ {p'}_B^- \nonumber\\ 
\hat s_{12} &=& 2 k_1\cdot k_2 = k_1^+ k_2^- + k_1^- k_2^+
- k_{1\perp} k_{2\perp}^* - k_{1\perp}^* k_{2\perp}\, ,\nonumber
\end{eqnarray}
however, anytime a difference of invariants is taken such that the leading
terms cancel, the invariants (\ref{crinv}) must be determined to
next-to-leading accuracy. The spinor products (\ref{spro}) become,
\begin{eqnarray}
\langle p_A p_B\rangle &\simeq& \langle p'_A p_B\rangle \simeq
-\sqrt{{p'}_A^+ {p'}_B^-} \nonumber\\
\langle p_A p'_B\rangle &\simeq& \langle p'_A p'_B\rangle =
-\sqrt{{p'}_A^+\over {p'}_B^+}\, p'_{B\perp} \nonumber\\
\langle p_A k_i\rangle &\simeq& \langle p'_A k_i\rangle =
-\sqrt{{p'}_A^+\over k_i^+}\, k_{i\perp} \nonumber\\
\langle k_i p_B\rangle &\simeq& -\sqrt{k_i^+ {p'}_B^-} \label{crpro}\\ 
\langle k_i p'_B\rangle &\simeq& -\sqrt{k_i^+\over {p'}_B^+}\,
p'_{B\perp} \nonumber\\
\langle p_A p'_A\rangle &\simeq& - p'_{A\perp} \nonumber\\
\langle p'_B p_B\rangle &\simeq& - |p'_{B\perp}| \nonumber\\
\langle k_1 k_2\rangle &=& k_{1\perp}\sqrt{k_2^+\over k_1^+} - 
k_{2\perp}\sqrt{k_1^+\over k_2^+}\, .\nonumber
\end{eqnarray}

\end{document}